\documentclass[twocolumn,numberedappendix,iop]{emulateapj}
\shorttitle{The Correlation Between Metallicity and Debris Disk Mass}
\shortauthors{G\'asp\'ar et al.}
\submitted{\today}
\journalinfo{The Astrophysical Journal}

\usepackage{apjfonts}
\usepackage{amsmath}
\usepackage{multirow}
\usepackage{wasysym}

\begin{document}

\title{The Correlation Between Metallicity and Debris Disk mass}

\author{Andr\'as G\'asp\'ar$^1$} 
\author{George H.~Rieke$^1$}
\author{Nicholas Ballering$^1$}
\affil{1) Steward Observatory, University of Arizona, Tucson, AZ, 85721\\
agaspar@as.arizona.edu, grieke@as.arizona.edu, ballerin@email.arizona.edu}

\begin{abstract}
We find that the initial dust masses in planetary debris disks are correlated with the metallicities 
of their central stars. We compiled a large sample of systems, including {\it Spitzer}, the 
Herschel DUNES and DEBRIS surveys, and {\it WISE} debris disk candidates. We also merged 33 
metallicity catalogs to provide homogeneous [Fe/H] and $\sigma_{\rm [Fe/H]}$ values. We analyzed this 
merged sample, including 222 detected disks (74 warm and 148 cold) around a total of 187 systems 
(some with multiple components) and 440 disks with only upper limits (125 warm and 315 cold), 
around a total of 360 systems. The disk dust masses at a common early evolutionary point in 
time were determined using our numerical disk evolutionary code, evolving a unique model for each of the 
662 disks backward to an age of 1 Myr. We find that disk-bearing stars seldom have metallicities less than 
${\rm [Fe/H]} = -0.2$ and that the distribution of warm component masses lacks examples with large mass around 
stars of low metallicity (${\rm [Fe/H]} < -0.085$). Previous efforts to find a correlation have been largely 
unsuccessful; the primary improvements supporting our result are: 1.) basing the study on dust masses, 
not just infrared excess detections; 2.) including upper limits on dust mass in a quantitative way; 
3.) accounting for the evolution of debris disk excesses as systems age; 4.) accounting fully for the 
range of uncertainties in metallicity measurements; and 5.) having a statistically 
large enough sample.

\end{abstract}

\section{Introduction}

The past three decades have seen the discovery of hundreds of planetary debris disks 
\citep[e.g.,][]{wyatt08} and a multitude of planets, either via radial-velocity surveys 
\citep[e.g., HARPS - ][]{mayor09}, transit surveys \citep[e.g., Kepler - ][]{borucki10}, 
microlensing \citep[e.g.,][]{bennett06}, or direct imaging \citep[e.g.,][]{marois08}. 
Together these discoveries have promoted exoplanetary system astronomy to a very prominent field 
of study. However, each of these tools probes only a small part of exoplanetary 
system behavior, and to understand more we need to combine these individual insights 
\citep[e.g.,][]{wright13}. 
 
Circumstellar planetary debris disks \citep[e.g.,][]{smith84,backman93} have the potential 
to probe: 1.) the evolution of the populations of small bodies, e.g. exoasteroids and 
exo-Kuiper-Belt-objects; 2.) recent major stochastic events such as exoasteroid collisions; 
3.) placement of critical zones, such as ice lines; and 4.) indirect indications of the 
existence and placement of planets. 

For example, we see debris disks fade at rates consistent with monotonic evolution from a formative 
stage, consistent with expectations for collisional cascades within exo-asteroidal belts, similar to the 
behavior of our asteroids \citep{wyatt07,gaspar13}. In some cases, particularly in young systems, there is 
direct evidence for huge collisions, either through spectral features indicative of fine silica dust 
\citep[e.g.,][]{johnson12b} or other very finely divided crystalline material, or through variability
\citep[e.g.,][]{meng14}. Imaged disks often show two distinct planetesimal zones analogous to the asteroid and
Kuiper Belts \citep[e.g.,][]{acke12,su13} and many unresolved systems have temperature structures suggestive of a similar
dual disk structure \citep[e.g.,][]{morales11,ballering13,chen14}. These structures may reflect ice line positions 
in the ancestral protoplanetary disks, with subsequent modification of the disk structure by planets. 
In fact, the putative planets have been imaged  in some debris disk systems \citep[Fomalhaut, HR 8799, $\beta$ Pic, 
HD95086, FW Tau, ROXs 12, ROXs 42B, HD 106906 ][]{kalas08,marois08,lagrange10,moor13,kraus14,currie14,bailey14}. 
The architectures of the disks in these systems are generally in agreement with the predictions from current 
disk -- planet interaction models \citep[e.g.][]{chiang09, rodigas14}. 
 
The influence of metallicity as measured by [Fe/H] can in principle provide additional insights to planetary system formation and evolution,
especially as an indicator of other heavy elements, such as the building blocks of organic molecules and planets (C, N, O, Si). 
As an example, the discovery that the number of massive, hot planets increases rapidly with increasing metallicity 
\citep[e.g.,][]{fischer05,johnson10}, in qualitative agreement with expectations, has inspired many hypotheses for 
the exact mechanism responsible \citep[e.g.,][]{johansen09,garaud11,johnson12,cossou14,nayakshin15}. 
How varying opacity affects - if at all - planet formation is still debated in planet formation theory.
While higher metal content is an indication of a generic higher density of available building material in
the systems, it also may play a crucial role in the physics driving planet formation itself.
While \cite{boss02} found only negligible variations in his gravitational instability models, \cite{cai06}, \cite{meru10}, 
and \cite{gammie01} find that lower metallicity systems cool faster and fragment more, aiding core accretion processes. 
This, however, also means that proto-planets in metal rich systems may have more time to accrete matter and form giant planets.

On first principles, higher metallicity should also favor the development of massive debris disks, since  
the necessary raw materials are then more abundant. Indeed, if the numbers of debris parent bodies 
are proportional to the metallicity, the prominence of debris disks might increase faster than proportionately since 
the collision rates should go roughly as the square of the space density of parent bodies. These trends might be 
countered if the sinks for planetesimals - assimilation into planets, or ejection from the system - have a 
countervailing dependence on metallicity. However, the numbers of the planets in the most common mass 
range - super-Earths to Neptunes \citep{malhotra15} - have at most a weak dependence on metallicity 
\citep[e.g.,][]{sousa11b, buchhave14,wang15a}. Thus, as direct sinks for planetesimal raw material they 
should have a neutral effect, or perhaps even reduce the available material at low metallicity. The fraction of stars 
with Jupiter-mass planets is low, $\lesssim$ 1\% \citep[e.g.,][]{wang15b, malhotra15}, so it would be surprising if their 
prowess at ejecting planetesimals had a noticeable effect on the overall incidence of debris disks. 

Despite these arguments, virtually all searches for a relation between stellar metallicity and the presence of debris 
disks \citep[e.g.,][]{bryden06,beichman06,greaves06,trilling08,marshall14,thureau14} find none. 
\citet{maldonado12} report a hint of a deficit of disks at low metallicity, but this behavior was not 
confirmed in a later paper \citep{maldonado15}, possibly because of the smaller sample in the latter work. Within the 
errors, these works are consistent with the incidence of debris disks being independent of 
the stellar metallicity\footnote{\citet{marshall14} even find preliminary evidence for an inverse correlation.}. 

With the exception of those by \citet{maldonado12,maldonado15}, these previous searches for a 
metallicity/debris disk relation have each been based on no more than $\sim$ 40 detected debris disk systens. 
Any systematic effects must be recognized out of a huge range of disk-creating 
activity, disk evolutionary behavior, and significant measurement errors, 
 placing a premium on averaging down the scatter with large numbers and on minimizing systematic effects. 
The era of major space-based surveys for debris disks is now past (with the demise of IRAS, ISO, cryo-Spitzer, Akari, 
cryo-WISE, and Herschel). It is therefore timely to combine the results from these missions to gain significantly in the 
statistical significance of debris disk studies. The study reported here is based on a combined catalog of 662 systems, 
including 222 detected examples and 440 meaningful upper limits. 

\setcounter{table}{0}
\begin{deluxetable}{ll|rrrrrr}[!t]
\tabletypesize{\scriptsize}
\tablecolumns{8}
\tablewidth{0pt}
\tablecaption{Properties and sources of the targets analyzed in the study. Stellar parameters were determined
via model spectra fitting. Disk catalogs are coded as follows: {\bf DU}NES, {\bf DE}BRIS, {\bf K.\ S}u et al. (2006), 
{\bf N.\ B}allering et al. (2013), {\bf J.\ S}ierchio et al. (2013), and {\bf W}ISE. Only the first 10 lines are displayed,
the full table is available online at ApJ or at CDS.\label{tab:feh}}
\tablehead{
\multicolumn{2}{c}{Name}     		& \colhead{Catalog} & \colhead{d}    & \colhead{$M_{\ast}$}    & \colhead{$L_{\ast}$}    & \colhead{$R_{\ast}$}    & \colhead{Age}    \\
\colhead{HIP} & \colhead{HD} 		& \colhead{}        & \colhead{(pc)} & \colhead{($M_{\odot}$)} & \colhead{($L_{\odot}$)} & \colhead{($R_{\odot}$)} & \colhead{(Gyr)} }
\startdata
000490	&	000105	&	JS,NB	&	39.4	&	1.13	&	1.29	&	1.06	&	0.170\\
000544	&	000166	&	DU,NB,W	&	13.7	&	0.98	&	0.60	&	0.94	&	0.240\\
000560	&	000203	&	NB	&	39.4	&	1.45	&	4.20	&	1.51	&	0.012\\
000682	&	000377	&	JS,NB	&	39.1	&	1.11	&	1.17	&	1.09	&	0.220\\
000910	&	000693	&	DU	&	18.7	&	1.39	&	3.07	&	1.50	&	3.000\\
000950	&	000739	&	DE	&	21.3	&	1.39	&	3.05	&	1.38	&	2.150\\
001031	&	000870	&	JS,NB	&	20.2	&	0.93	&	0.48	&	0.84	&	2.300\\
001134	&	000984	&	JS	&	47.1	&	1.27	&	2.14	&	1.25	&	0.250\\
001292	&	001237	&	DE	&	17.5	&	0.99	&	0.64	&	0.88	&	0.300\\
001368	&	-	&	DE,NB	&	15.0	&	0.65	&	0.11	&	0.68	&	0.900
\enddata
\end{deluxetable}

We also introduce new approaches to reduce systematic effects. To be specific, the previous studies simply 
compared the stellar metallicity distribution of sources with or without detected excesses of any size, 
whereas it is the disk masses that should correlate with metallicity. Furthermore, the previous work does not take debris 
disk evolution into account which, if ignored, will dilute any other effects on disk incidence because after 
a certain age stars of all metallicities will have few debris disks. Thus, a more sensitive search for a 
correlation between metallicity and the presence of debris disks should convert the debris disk emission 
levels to mass, correct to a common time, preferably one at a young age where the result should be representative of 
the dust mass in the protoplanetary disk \citep{wyatt07,gaspar13}, and work with the full distribution of detected 
masses and upper limits. It would also be desirable to treat separately the warm and cold debris
disk components \citep{morales11,kennedy14} that evolve independently and at quite different rates \citep{gaspar13}. 
In this paper, we introduce all of these improvements. We also review the previous assumptions (e.g., on metallicity 
determination vs.\ spectral type) to identify any that might have undermined detection of an effect. Because of the new
modeling approach and the large sample, we do find a dependence of disk mass on metallicity for both warm and cold disk 
components, at modest statistical significance.

We present this work as follows. In section \ref{sec:sample}, we detail the observational sample 
we use for our study, while in section \ref{sec:feh} we describe the methods used to estimate the metallicities of 
the sources in the sample. The modeling methods are elaborated in section \ref{sec:zero}, 
while in section \ref{sec:results} we discuss the results. Section \ref{sec:conclude} summarizes our findings.

\section{The disk sample}
\label{sec:sample}

To build the largest possible sample of debris disks, we merged data from multiple sources. We separate 
the final sample into two groups: warm and cold debris disks. Systems with {\it Spitzer} 24 $\micron$ excesses
or upper limits or with WISE W4 excesses are included in the warm disk sample, while systems with
{\it Spitzer} 70 and/or {\it Herschel} 70/100 $\micron$ data (either excesses or upper limits) are
included in the cold disk sample. 

In the following subsections we detail each dataset used to compile our final warm and cold disk samples.
For the study of warm disk components, based on our results in \cite{gaspar13}, we selected only
systems up to $0.5~{\rm Gyr}$ in age, as detectable systems older than this limit are very likely to 
have undergone a recent collisional event. We also analyze the cold sample including only sources 
up to 5 Gyr in age, due to similar evolutionary effects recognized in \cite{sierchio14}.
For sources where disks were not detected at either 24 $\micron$ (MIPS/{\it Spitzer} data) or at 
70/100 $\micron$ (MIPS/{\it Spitzer} or PACS/{\it Herschel} data), we used the upper limits given by 
the observations. From the {\it ROSAT} and {\it WISE} selected sample we only used the 
detections\footnote{The {\it WISE} sample would have yielded a huge number of sources to run upper limit 
models for, and would not have contributed significantly to our results because these limits are generally less 
stringent than those from {\it Spitzer}.}. Our final disk catalog is shown in Table \ref{tab:feh},
with stellar mass, luminosity, and radius estimates from spectral model fitting.

\subsection{The {\it Spitzer} IRS Selected Debris Disks}
\label{sec:IRS}

\cite{ballering13} analyzed {\it Spitzer} IRS spectra and MIPS 
24 and 70 $\micron$ data of 546 main sequence stars, characterizing 170 cold and 117 
warm disk components of 214 sources with excess. A similar analysis of the IRS sample was performed 
by \cite{chen14}. \cite{ballering13} specifically categorized debris disk systems as having either 
a warm and/or cold component and specific fluxes for each component are given, therefore, sources 
from the \cite{ballering13} IRS sample are included using their definitions and fluxes. We refer the 
reader to \cite{ballering13} for the details on how the temperatures of the components and the ages 
of the systems were determined. We required the sources in our study to have defined ages and 
metallicities, which narrowed the \cite{ballering13} sample to 140 systems with excesses (48 warm 
and 112 cold components and 36 and 28 upper limits, respectively)\footnote{We also removed the HIP 106741
system from the \cite{ballering13} sample, due to it likely being background contaminated \citep{panic13}.}.

\setcounter{table}{1}
\tabletypesize{\scriptsize}
\begin{deluxetable*}{llrrl}[!t]
\tablecolumns{5}
\tablewidth{0pt}
\tablecaption{The metallicity catalogs combined in our work listed in the order of merging.
Linear conversion factors (${\rm [Fe/H]}_{\rm merged} = a + b {\rm [Fe/H]}_{\rm original}$) 
are also given.\label{tab:FeH}}
\tablehead{\colhead{Sequence} & \colhead{Catalog} & \colhead{$a$} & \colhead{b} & \colhead{Notes}}
\startdata
1       &       \cite{valenti05}        &       0.000   &       1.000   &       Base system in this study; spectroscopic; has common errors of $\sigma_{\rm [Fe/H]}=0.03$\\
2       &       \cite{wu11}             &       0.027   &       1.033   &       1273 stars; spectroscopic\\
3       &       \cite{takeda05}         &       0.003   &       0.966   &       160 mid-F to early K stars, spectroscopic\\
4       &       HARPS \citep{sousa08,sousa11a,sousa11b} & 0.000 &       1.000   & Spectroscopic, common sources were adopted from the latest papers \\
5       &       \cite{robinson07}       &       -0.009  &       0.885   &       N2K low-res spectroscopic survey, 1907 stars, $\sigma_{\rm [Fe/H]}=0.07$ used for all sources\\
6       &       Geneva-Copenhagen Survey \citep{casagrande11} & -0.001 & 0.967 & Latest, updated values of the spectroscopic survey data \\
7       &       \cite{suchkov03}        &       0.040   &       0.760   &       [Fe/H] calculated from Str\"omgren photometry, based on \cite{schuster89}\\
8       &       NStars survey \citep{gray03,gray06}     &       0.045   &       1.030   & NStars spectroscopic survey of stars earlier than M0 and closer than 40 pc \\
9       &       \cite{maldonado12}      &       0.015   &       0.910   &       Only new spectroscopic data from Table 5 merged \\
10      &       \cite{jenkins08}        &       0.005   &       0.964   &       High-res spectroscopic survey of 353 solar-type stars\\
11      &       \cite{koleva12}         &       -0.002  &       0.939   &       Fitting of the New Generation Stellar Library (STIS/HST)\\
12      &       \cite{bond06}           &       0.091   &       1.330   &       High-res spectroscopic study of 136 G-type stars\\
13      &       \cite{erspamer03}       &       0.027   &       0.970   &       High-res spectroscopic study of 140 A-F type stars\\
14      &       \cite{saffe08}          &       0.070   &       0.850   &       Spectroscopic study of 113 BAFGK-type Southern stars\\
15      &       \cite{takeda09}         &       0.040   &       0.950   &       Spectroscopic study of 120 A-type stars\\
16      &       \cite{gerbaldi07}       &       -0.020  &       0.660   &       [Fe/H] calculated from Str\"omgren photometry\\
17      &       \cite{gebran10}         &       0.000   &       1.000   &       Spectroscopic study of 44 A and F-type Hyades stars\\
18      &       \cite{katz11}           &       -0.015  &       0.915   &       Spectroscopic study of 400 stars\\
19      &       \cite{paulson06}        &       0.015   &       1.760   &       Spectroscopic study of $\beta$ Pic MG stars, errors of $\sigma_{\rm [M/H]}=0.08$ used\\
20      &       \cite{paunzen02}        &       -0.190  &       0.893   &       Spectroscopic study of $\lambda$ Bo\"otis stars\\
21      &       \cite{kunzli98}         &       0.080   &       1.520   &       [M/H] values adopted, with generic $\sigma_{\rm [M/H]}=0.2$\\
22      &       \cite{philip80}         &       -0.056  &       0.691   &       [Fe/H] from Str\"omgren photometry, with generic errors of $\sigma_{\rm [M/H]}=0.15^{\dagger}$\\
23      &       \cite{gebran08b}        &       0.070   &       1.000   &       High-res spectroscopic study of the Pleiades.\\
24      &       \cite{jenkins09}        &       0.000   &       1.000   &       Not enough data to determine conversion \\
25      &       \cite{andrievsky98}     &       0.000   &       1.000   &       Not enough data to determine conversion \\
26      &       \cite{guillout09}       &       0.000   &       1.000   &       Not enough data to determine conversion \\
27      &       \cite{lemke89}          &       0.000   &       1.000   &       Not enough data to determine conversion \\
28      &       \cite{gebran08a}        &       0.000   &       1.000   &       Not enough data to determine conversion \\
29      &       \cite{fuhrmann08}       &       0.016   &       0.938   &       Volume limited spectroscopic study of F-K stars\\
30      &       \cite{favata97}         &       -0.005   &      0.957   &       Spectroscopic study of 91 G and K stars\\
31      &       \cite{luck05}    	&       0.004   &       0.892   &       Spectroscopic study of 114 stars within 15 pc\\
32      &       \cite{luck06}    	&       -0.008   &      0.831   &       Spectroscopic study of 216 stars within 15 pc\\
33      &       \cite{ammons06}         &       -0.100  &       1.000   &       Only remaining disk sources added that were not available from other catalogs
\enddata
\tablenotetext{$\dagger$}{Metallicities had to be recalculated based on the procedure given in the paper, as on-line data
had errors of missing $\pm$ indexes of metallicities.}
\end{deluxetable*}

\subsection{The \cite{gaspar13} Sample}
\label{sec:gaspar}

In \citet{gaspar13}, we analyzed the {\it Herschel} DEBRIS and DUNES
surveys \citep{eiroa13,thureau14,moro-martin15}, and supplemented the {\it Herschel} results with 
{\it Spitzer} MIPS 24 and 70 $\micron$ data. The \cite{gaspar13} results also included MIPS 24 and 70 
$\micron$ photometry from \cite{sierchio14}
and \cite{su05}, which we include in our current sample as well; however, we do not
include stellar open cluster data. We took ages from \cite{gaspar13} and \cite{sierchio14} but 
updated the {\it Herschel} flux measurements with those from \cite{eiroa13}, \cite{thureau14}, 
and \cite{moro-martin15}. Of the 387 systems included from the \cite{gaspar13} sample, 60 are also part of the 
\cite{ballering13} IRS sample; the radial distances of the disks around them 
and the warm and/or cold disk flux contributions to their excesses were included from the IRS 
analysis. Excluding the IRS sources, 327 were included from the \cite{gaspar13} sample that had 
age and metallicity values, of which 105 were up to 500 Myr old.

\subsection{The {\it ROSAT} and {\it WISE} selected Sample}

In section \ref{sec:feh}, we compile a list of metallicity catalogs that had data for at least one of the sources
either from the {\it IRS} or \cite{gaspar13} datasets. The combined non-redundant source count of these metallicity 
catalogs was 20811. Of the 20811 sources in the compiled catalog of sources with a metallicity value, essentially all 
have been measured with WISE and 1753 had ROSAT measurements. We used the latter to determine stellar ages, as 
in \citet{sierchio14}
\begin{equation}
\log_{10}\left(t/{\rm yr}\right) = 1.2-2.307\log_{10}\left(R_x\right)-0.1512\left[\log_{10}\left(R_x\right)\right]^2\;,
\end{equation}
where $R_x$ is the ratio of the X-ray to total luminosity:
\begin{equation}
R_x = 10^{20}\frac{L_x}{L}\;.
\end{equation}
This age estimate is accurate to within 0.1-0.2 dex \citep{mamajek08,sierchio14}.
X-ray fluxes were calculated from HR1 {\it ROSAT} data following \cite{schmitt95},
while total luminosities were estimated from SED fits. A total of 1736 sources had ROSAT, WISE,
and metallicity data. The distribution of the W3-W4 colors peaked at 0.054 mag, with a rms scatter of 0.042 mag. 
Any source in the sample that had $W3-W4\ge0.181$ (i.e., a nominal 3-$\sigma$ detection) was further analyzed. 
If its excess at W4 was above the predicted levels by $4 \sigma_{W4}$ (using its individual measured error) and 
if its SED fit passed a by-eye examination for consistency, we added the source to our study. Only 59 sources 
passed all of these criteria. Of these, only 40 were younger than 0.5 Gyr; of the remaining 19, 5 were identified in other
catalogs and further selected for the cold disk studies (as either detections or upper limits).

Similar WISE catalog searches were performed by \cite{patel14}, \cite{wu13}, and \cite{cruz14}.
\cite{patel14} searched for WISE excess (W3 and W4) sources within 75 pc, placing special emphasis on finding sources
with saturated photometry. \cite{wu13} searched for excesses at W4 for Hipparcos sources
within 200 pc, while \cite{cruz14} placed a photometry limit of $V=15^{\rm mag}$ on their WISE catalog search.
Of the 59 sources we identified, 57 are within 200 pc (compared to the 103 new detections by \cite{wu13}),
41 are within 75 pc (compared to the 106 new detections by \cite{patel14}), and all are brighter than V=10.12 mag
(well above the \cite{cruz14} cutoff). Of the 59 sources we initially identified 27 were also found by either the
\cite{wu13} or the \cite{patel14} analysis (or by both), however, 32 are possibly new sources. \cite{patel14}
go through a rigorous set of 15 criteria that their data have to meet to be considered an excess source. To ensure
that our data meet their criteria as well, we removed the new sources from our analysis as well as the
sources that were older than 0.5 Gyr and only identified in the {\it ROSAT/WISE} sample. This finally leaves
only 29 sources in this sample, of which 20 are younger than 0.5 Gyr. Of these 20, 11 are exclusively
identified in the WISE study. This sequence of steps demonstrates that our sample is not missing a large number
of strong WISE detections.

\setcounter{table}{2}
\begin{deluxetable}{lll|rr|l}[!t]
\tablecolumns{6}
\tablewidth{0.45\textwidth}
\tablecaption{
The combined metallicity catalog of 20811 sources. The reference numbers are based on the sequence numbers in
Table \ref{tab:FeH}. Only the first 10 lines are displayed, the full table is available online at ApJ or at CDS.
\label{tab:mets}}
\tablehead{
\colhead{HIP} & \colhead{HD} & \colhead{Simbad} & \colhead{[Fe/H]} & \colhead{$\sigma_{\rm[Fe/H]}$} & \colhead{Refs}}
\startdata
HIP000004       &       HD224707        &       HD 224707       &       -0.45   &       0.10    &       6       \\
HIP000020       &       HD224723        &       HD 224723       &       -0.25   &       0.10    &       6,33    \\
HIP000023       &       HD224742        &       HD 224742       &       -0.26   &       0.08    &       6,7,33  \\
HIP000033       &       HD224743        &       HD 224743       &       -0.06   &       0.10    &       6,33    \\
HIP000034       &       HD224758        &       HR 9078 	&       0.05    &       0.07    &       6,7,22,33       \\
HIP000038       &       HD224752        &       HD 224752       &       -0.14   &       0.10    &       6,33    \\
HIP000039       &       HD224763        &       NLTT 58719      &       -0.16   &       0.10    &       6,33    \\
HIP000042       &       HD224771        &       HD 224771       &       -0.02   &       0.10    &       6,33    \\
HIP000050       &       HD224782        &       HD 224782       &       0.44    &       0.10    &       6,33    \\
HIP000055       &       HD224783        &       HD 224783       &       0.16    &       0.10    &       6,33
\enddata
\end{deluxetable}

\section{Metallicities}
\label{sec:feh}

To determine the metallicities of the sources, we first compiled a list of all catalogs that had data for 
at least one of our stars. Some catalogs themselves were compilations 
(e.g.\ the PASTEL catalog - \citeauthor{soubiran10}\ \citeyear{soubiran10} or 
\citeauthor{taylor05} \citeyear{taylor05}), while a few contained sources republished multiple times
\citep[e.g.\ the HARPS papers,][]{sousa08,sousa11a,sousa11b} or superseded/reevaluated data 
\citep[e.g.\ the Geneva-Copenhagen Survey,][]{nordstrom04,holmberg09,casagrande11}. We made sure 
to include data from each observation only once - the latest - in these cases, or in the case of 
the compilation catalogs to use them as aids in finding additional catalogs that were missing from Vizier. 
Our final database is assembled using 33 catalogs.

The metallicity values determined in each spectroscopic survey depend on instrument calibration, the lines
used, whether [Fe/H] or total metallicity [M/H] is determined, and the synthetic model atmospheres that
the observations are compared to. Because of these, catalogs may experience a metallicity-dependent 
scaling offset. To correct for this, before combining the catalogs, we converted the values from each 
catalog into a ``common system'', which we chose to be that of \cite{valenti05}. The conversion 
was performed by simple linear regression between the data sets, as 
${\rm [Fe/H]}_{\rm merged} = a + b {\rm [Fe/H}]_{\rm original}$. We merged the catalogs 
together sequentially, in the order of the number of common sources between them and the sequentially merged 
catalog, recalculating the errors in metallicity as the rss of the errors in each catalog and the average 
of the metallicities weighted by their errors. The conversion factors
were calculated preceding each sequential merging to ensure a larger number of points in the
linear regression. 

In Table \ref{tab:FeH}, we list the metallicity catalogs we merged in sequential
order and the conversion factors used. We also note special circumstances with some of the data.
Our final merged catalog contains [Fe/H] and $\sigma_{\rm [Fe/H]}$ values for 20811
sources, all in a common system. The sample includes 48 chemically peculiar $\lambda$ Bo\"otis type stars
for which the metallicities are not indicative of the environment they formed in, but of the peculiar 
surface composition of the star. None of these sources were analyzed in our work, even if they
had a disk or upper limit measurement. We present our combined metallicity catalog 
in Table \ref{tab:mets} for future reference.

Outlier metallicities may result from unidentified abundance peculiarities. 
Therefore, for our analysis we only considered sources with metallicities between 
$-0.6 \ge {\rm [Fe/H]} \le 0.4 $, based on the results of \cite{casagrande11} that demonstrate 
that virtually all stars in the solar neighborhood have metallicities within this range. Our original sample 
included 28 such sources. We also removed 20 sources with ${\rm v} \sin\left(\iota\right)$ values above 
200 km s$^{-1}$, as such large rotational velocities make it difficult to determine the stellar 
metallicity accurately (R.\ Gray, private communication).

One important issue is that the measured stellar abundances may not be representative of the average through the 
entire stellar volume. Stars without significant envelope convection tend to have strata with varying metal 
content \citep[e.g.,][]{leblanc09}, that can yield an apparent metallicity that differs significantly from 
the average for that star. The lack of convective mixing can lead to other types of metallicity anomaly, such 
as the $\lambda$ Bo\"otis phenomenon, perhaps produced by accretion of relatively small amounts of material onto 
the stellar surface layer \citep[e.g.,][]{jura15}. Presumably similar processes at a lower level can produce 
anomalies below the threshold for identification as a {\it bona fide} $\lambda$ Bo\"otis-type star. In fact, a 
significantly larger scatter in A-star metallicity measurements relative to those for later-types is seen in open 
clusters \citep{gebran08a,gebran08b,gebran10}. Debris disk samples generally contain a number of A-stars, in which case 
the measured abundances against which the disk presence is measured may have large errors and systematic biases. 
Therefore, we have looked at the possible correlations both for our entire sample, and just for those stars cool 
enough to have substantial convective outer layers that should produce more homogeneous metallicity behavior. To 
be conservative, we have divided the nonconvective and convective samples at the A9/F0 transition\citep{bohm66,simon02}. 
For similar reasons, metallicity values of early-type stars in nearby moving groups were averaged with the moving 
group average value. 

Another consideration is that debris at detectable levels is largely absent for M-stars \citep[e.g.,][]{heng13}, 
although the incidence of excesses seems to not change significantly down through early K-stars \citep{sierchio14}. 
Therefore, to avoid contaminating any effects with stars wth only upper limits regardless of metallicity, we have 
cut from the sample stars later than K5. The sample trimmed of A-stars and late-type ones is reduced by 
127 sources (20 warm detections, 41 warm upper limits, 103 cold detections, 75 cold upper limits). Many of our 
analyses will be done both for this trimmed sample and for the full sample; we find relatively little difference in 
the results. 

\section{Zero Aging the Systems}
\label{sec:zero}

Debris Disks are in a persistent state of collisional evolution,
with the smaller particles eroding away at the larger rocks, producing more dust and more erosion \citep{wyatt08}.
A quasi steady state is reached \citep{thebault03,krivov05,gaspar12a}, without runaway small dust production, 
due to the blowout of the smallest $\micron$ size particles in the systems via radiative and corpuscular forces 
\citep{burns79}. Due to the continuous collisional erosion, over time, the disks lose mass and their thermal emission in the
mid and/or far-IR fades away \citep{wyatt08,gaspar13}. Initially, this process is dominated by evolution from initial
conditions set early in the life of the star, probably even by the mass of its protoplanetary disk \citep{wyatt07,gaspar13}. 
The influence of large collisions is transitory and generally minor \citep{kenyon05} until the primary evolutionary phase 
is past, when the incidence of disks drops precipitously \citep{sierchio14}. Many of the remaining disks are probably 
the result of recent and transitory events \citep{gaspar13}. This behavior provides an opportunity to take a disk of 
any age within the slowly-decaying initial phase and use models of its evolution to estimate its characteristics at a 
young age. That is, it allows de-aging debris disks for an ensemble of stars of different ages to a common age. In 
this paper, we use our evolutionary collisional cascade model, CODE-M \citep{gaspar12a}, to carry out these calculations.

\subsection{Procedure}

We model the collisional evolution of each individual disk using the collisional cascade code. Apart from the disk 
radial distance, disk width, disk height, the spectral/emission properties of the host star, and disk mass, all other 
variables of our models are the same as that of our Reference Model in \cite{gaspar12b}. The radial distance was 
determined from the thermal location of the disks, assuming that the emitting dust is in thermal equilibrium and that 
the stellar radiation is well approximated by the best fitting Kurucz model to the UBVRIJHK photometric data of each 
source. The fitting was done in logarithmic flux space, following filter bandpass corrections. Relatively accurate 
thermal locations are available for disks observed with {\it Spitzer} IRS; for all other sources we assumed a standard
warm and cold disk thermal location of 190 and 60 K, respectively \citep{morales11,ballering13}. The width and
height of the disks were set to 10\% of the radial distance, while the radiation forces acting on the particles were 
calculated from the best fitting Kurucz models, assuming astronomical silicate particles \citep{draine84}. Additional 
numerical and modeling considerations are described in \cite{gaspar12a}. For the current study, the relative values of 
the disk parameters are important, so a consistent set of models is adequate without requiring that the models be 
correct in an absolute sense (which would, e.g., require consideration of more complex grain compositions). 

\begin{figure*}[!t]
\begin{center}
\includegraphics[angle=0,scale=0.7]{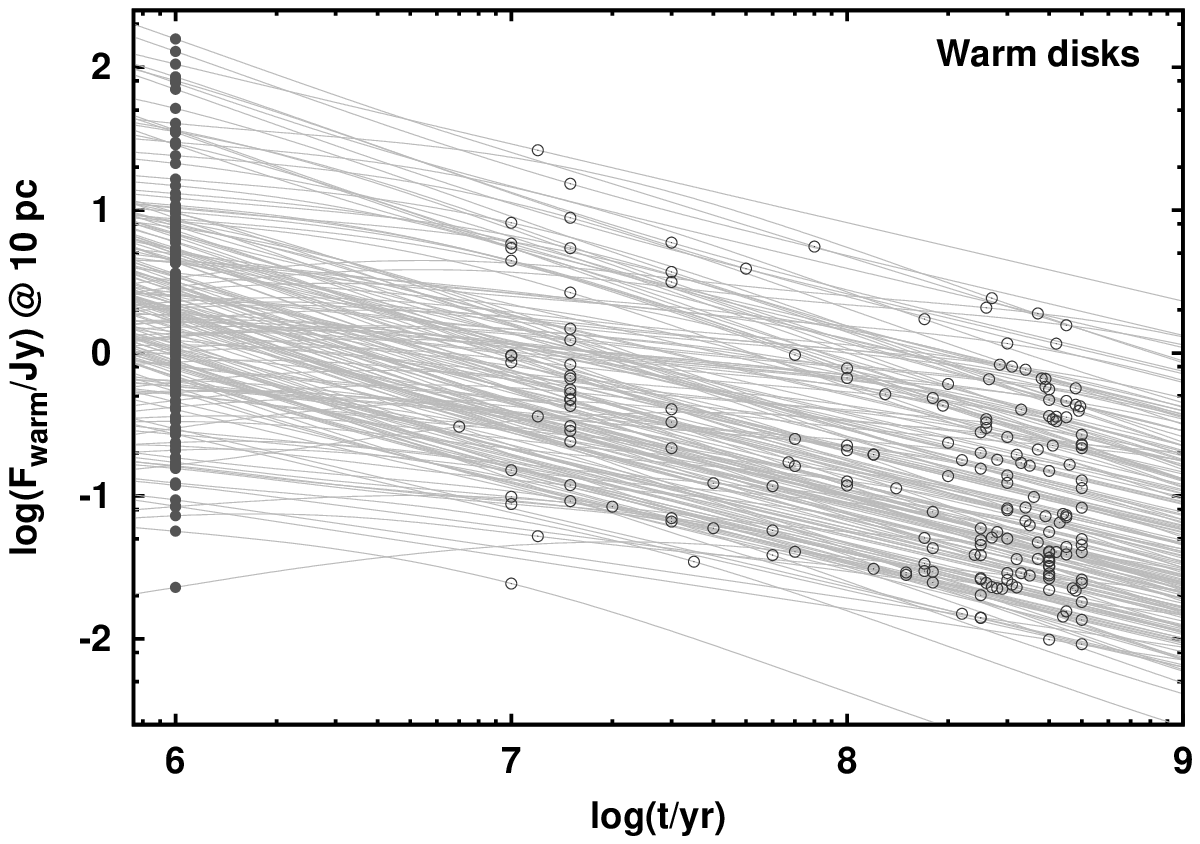}
\includegraphics[angle=0,scale=0.7]{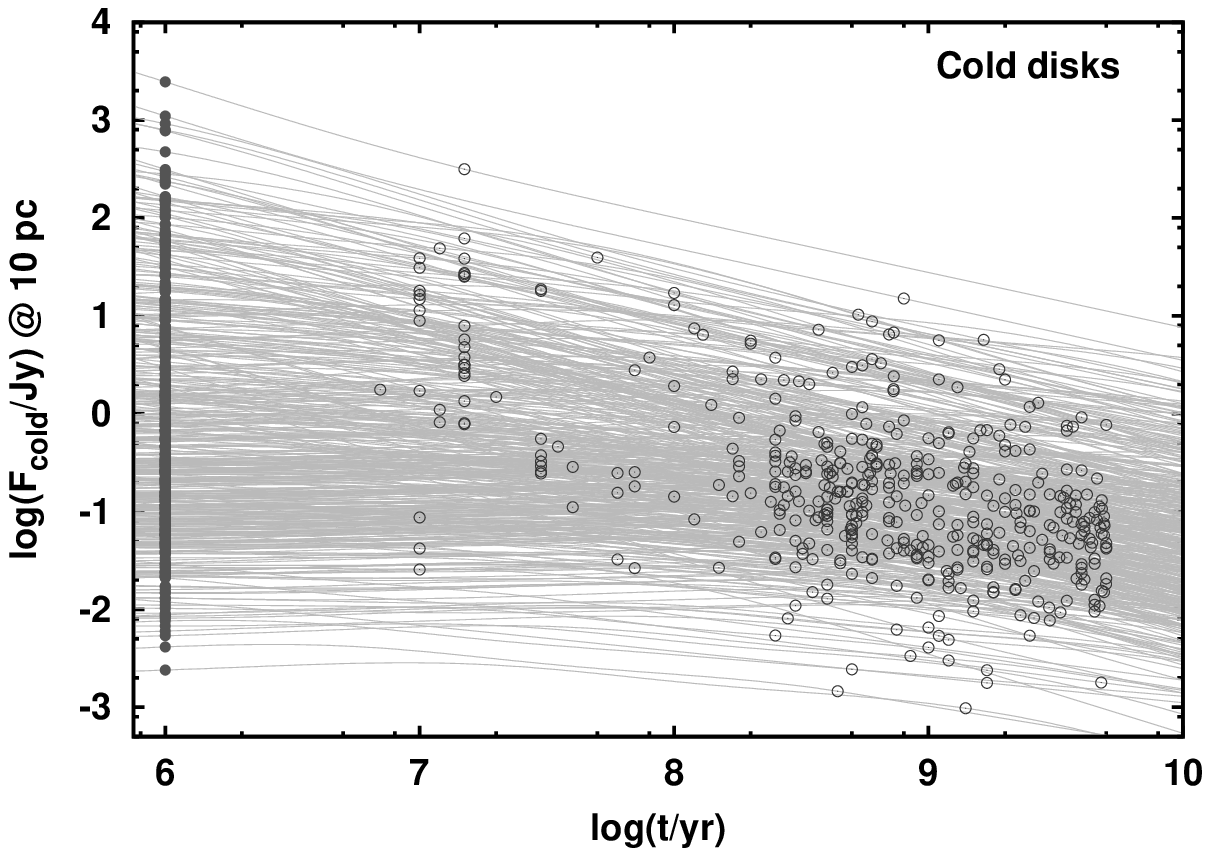}
\caption{The process of de-aging the disk fluxes is shown for the warm disk sample ({\it left panel}) and cold disk
sample ({\it right panel}). Since the disk and stellar parameters vary from one system to another, each evolution
curve had to be calculated for each individual system separately. Once the mass scaling of each curve is found that 
allows the curve to intersect the observed photometry, the mass can be traced to any point in the evolution. The observed
photometry points are shown with empty circles, while the traced points at 1 Myr are shown with filled circles.
The disk fluxes (photosphere subtracted) are scaled to a hypothetical source distance of 10 pc.}
\label{fig:deage}
\end{center}
\end{figure*}

The modeled evolution of the system flux is fitted to the observed emission using the scaling laws we introduced 
in \cite{gaspar13}. As detailed there, since the evolution is always slower than $t^{-1}$ and the evolutionary 
tracks scale along $t^{-1}$ as a function of system mass, there is only a single solution to the scaling of the 
evolutionary track. This scaling yields the true zero-age system dust mass in units of the original modeled 
system mass. Once the correct scaling is found, system variables (such as dust mass) can be calculated at any 
point in time during the evolution of the system.

Modeling of the disks took on average $\approx$ 8 days for each disk, at which point the mid- and/or far-IR
flux evolutionary track of the system could be scaled along a $t^{-1}$ axis in such a way that it intersected 
the observed system flux, as detailed above. Using a 32 node computer cluster, running the code for all 662 
systems took altogether over 15 CPU years.

In Figure \ref{fig:deage}, we show the process of deaging the systems. Since our initial particle size
distribution is slightly shallower than the final steep size distribution that results from a collisional
cascade \citep{gaspar12b}, the emission of the disks actually increases in the initial phase of the 
disk evolution while the system is settling into its quasi steady state. This can be seen for some of the lower
initial mass systems in Figure \ref{fig:deage}. Therefore, we define ``zero-age'' at 1 Myr. The initial conditions of the 
systems are somewhat uncertain. However, apart from the previously mentioned settling phase, they will not greatly 
affect the evolution. We verified this by analyzing the results at multiple ages near the ``zero-age'' point. 

We calculated the total system dust mass at these points in time, by integrating the particle size 
distributions calculated by our numerical code at the corresponding model time and scaling by the 
factor determined from the flux fits for each system. The distributions
are integrated up to 1 mm in radius and the final disk dust masses are divided by the masses of their central stars
(in solar units) to place all systems on similar scales. This final step also makes the disk mass metric analogous 
to metallicity. Due to the many orders of magnitude covered in disk mass space and only a few factors of difference 
in stellar masses, the normalization by stellar mass did not make a noticeable difference in our final results.
In our Figures we plot ``reduced masses'', which are the disk dust masses divided by Earth mass and the mass
of the central star in solar units ($M_{\rm reduced} = M_{\rm dust}/M_{\oplus}/M_{\ast}/M_{\odot}$).
We summarized the warm and cold disk properties in Tables \ref{tab:warmd} and \ref{tab:coldd}, respectively.

\subsection{Application}

As shown in \cite{gaspar13}, the observed fraction of warm debris disks, with excesses 10\% above the stellar
photosphere, decay to only a few percent for sources older than $\sim 500~{\rm Myr}$. This trend is independent 
of the spectral-type of the host star. Since our code models quasi-steady state collisional cascades, we only 
include sources with ages up to 500 Myr in our correlation analysis, as many sources that are older are likely 
experiencing a late-stage stochastic event. With this cut, our warm disk sample has 199 sources, 188 observed by
MIPS/{\it Spitzer} at $24~\micron$ (63 detections and 125 upper limits) and 11 observed only by WISE at $22~\micron$ 
(only detections).

Unlike the warm disks, cold debris disks can evolve in a quasi-steady state collisional cascade
well up to 5 Gyr \citep{gaspar13,sierchio14}, therefore we made an age cut at 5 Gyr in the sample 
when analyzing a possible correlation between the cold disk dust mass and host star metallicity. 
Our cold disk sample has 463 sources (148 detections and 315 upper limits), observed by 
MIPS/{\it Spitzer} at $70~\micron$ and/or by PACS/{\it Herschel} at $100~\micron$. If a source was 
observed at both bands, we averaged the disk dust mass predicted for each band.

\setcounter{table}{3}
\begin{deluxetable*}{lr|rr|rr|rr|r}[!t]
\tablecolumns{9}
\tablewidth{0pt}
\tablecaption{
The observed and de-aged parameters of the warm disk sources at $t=1~{\rm Myr}$ and
at the actual ages. Dust masses are calculated for particles ranging in size from the blowout 
limit to 1 mm in radius. The fluxes displayed are for the disks only (photosphere subtracted)
and calculated for the system being @ 10 pc to allow for comparison. The type denotes
either 22 $\micron$ {\bf W}ISE detection, 24 $\micron$ {\bf S}pitzer detection, or 24 $\micron$ Spitzer {\bf U}pper limit.
Only the first 10 lines are displayed, the full table is available online at ApJ or at CDS. \label{tab:warmd}}
\tablehead{
\multicolumn{2}{c}{Name} & \colhead{$T_{\rm warm}$} & \colhead{$R_{\rm warm}$} & \colhead{$F_{\rm warm}$} & \colhead{$F_{\rm warm}(1)$} & \colhead{$M_{\rm dust}$} & \colhead{$M_{\rm dust}(1)$} & \colhead{Type}\\
\colhead{HIP} & \colhead{HD} & \colhead{(K)} & \colhead{(AU)} & \multicolumn{2}{c}{(mJy)@10 pc} & \multicolumn{2}{c}{($\log(M_{\rm dust}/M_{\oplus})$)} & \colhead{}}
\startdata
HIP000490       &       HD000105        &       190.00  &       2.44    &       50.8    &               1988.1  &             -6.369   &               -4.777  &               U\\
HIP000544       &       HD000166        &       126.40  &       3.79    &       38.6    &               1963.1  &             -5.920   &               -4.214  &               S\\
HIP000560       &       HD000203        &       190.00  &       4.40    &       52.1    &               72.5    &             -6.248   &               -6.092  &               U\\
HIP000682       &       HD000377        &       119.30  &       5.90    &       177.6   &               9525.3  &             -5.079   &               -3.350  &               S\\
HIP001134       &       HD000984        &       190.00  &       3.14    &       59.1    &               3392.0  &             -6.257   &               -4.499  &               U\\
HIP001292       &       HD001237        &       190.00  &       1.72    &       25.7    &               852.3   &             -6.725   &               -5.206  &               U\\
HIP001473       &       HD001404        &       132.90  &       20.30   &       461.7   &               1074.9  &             -4.285   &               -3.870  &               S\\
HIP001481       &       HD001466        &       216.60  &       1.96    &       235.1   &               4381.4  &             -5.881   &               -4.611  &               S\\
HIP001803       &       HD001835        &       190.00  &       2.19    &       74.7    &               2741.8  &             -6.215   &               -4.650  &               U\\
HIP002578       &       HD003003        &       194.40  &       8.13    &       3136.1  &               35296.6 &             -4.285   &               -3.234  &               S
\enddata
\end{deluxetable*}

\setcounter{table}{4}
\begin{deluxetable*}{lr|rr|rr|rr|rr|r}[!t]
\tablecolumns{11}
\tablewidth{0pt}
\tablecaption{
The observed and de-aged parameters of the cold disk sources at $t=1~{\rm Myr}$ and
at the actual ages. Dust masses are calculated for particles ranging in size from the blowout 
limit to 1 mm in radius. Where dust masses could be calculated from both 70 and 100 $\micron$ observations, the dust masses 
were averaged. The fluxes displayed are for the disks only (photosphere subtracted)
and calculated for the system being @ 10 pc to allow for comparison. The type denotes either a 
detection at 70 or 100 $\micron$ ({\bf 70D} or {\bf 100D}) or an upper limit ({\bf 70U} or {\bf 100U}). 
Only the first 10 lines are displayed, the full table is available online at ApJ or at CDS. \label{tab:coldd}
}
\tablehead{
\multicolumn{2}{c}{Name} & \colhead{$T_{\rm cold}$} & \colhead{$R_{\rm cold}$} & \colhead{$F_{70}$} & \colhead{$F_{70}(1)$} & \colhead{$F_{100}$} & \colhead{$F_{100}(1)$} & \colhead{$M_{\rm dust}$} & \colhead{$M_{\rm dust}(1)$} & \colhead{Type}\\
\colhead{HIP} & \colhead{HD} & \colhead{(K)} & \colhead{(AU)} & \multicolumn{2}{c}{(mJy)@10 pc} & \multicolumn{2}{c}{(mJy)@10 pc} & \multicolumn{2}{c}{($\log(M_{\rm dust}/M_{\oplus})$)} & \colhead{}}
\startdata
HIP000490 & HD000105 & 47.84  & 33.70 & 2263.236 & 11950.800 &     -   &       -  & -2.501  & -1.779  & 70D\\
HIP000544 & HD000166 & 49.92  & 21.38 &  125.698 & 256.054   & 87.705  & 160.267  & -3.907  & -3.620  & 70D,100D\\
HIP000560 & HD000203 & 128.70 & 9.63  & 1097.957 & 1983.503  &     -   &       -  & -3.962  & -3.705  & 70D\\
HIP000682 & HD000377 & 43.89  & 37.44 & 2253.957 & 11157.030 &     -   &       -  & -2.380  & -1.684  & 70D\\
HIP000910 & HD000693 & 60.00  & 34.62 &   38.627 & 58.293    & 14.727  &  20.442  & -4.688  & -4.531  & 70D,100D\\
HIP000950 & HD000739 & 60.00  & 34.50 &   44.455 & 69.758    & 161.114 & 534.236  & -4.133  & -3.766  & 70U,100U\\
HIP001031 & HD000870 & 50.16  & 18.93 &   68.318 & 4293.173  &     -   &       -  & -4.164  & -2.366  & 70D\\
HIP001134 & HD000984 & 60.00  & 28.93 &  371.446 & 742.581   &     -   &       -  & -3.567  & -3.262  & 70U\\
HIP001292 & HD001237 & 60.00  & 15.80 &   26.943 & 45.663    & 31.046  &  62.718  & -4.659  & -4.390  & 70U,100U\\
HIP001368 &        - & 36.84  & 15.55 &   33.272 & 113.487   & 87.006  & 368.355  & -3.824  & -3.244  & 70D,100D
\enddata
\end{deluxetable*}

\section{Results}
\label{sec:results}

\subsection{Confirming Previous Evidence for a Metallicity Dependence \citep{maldonado12}}

\citet{maldonado12} merged multiple sets of data on debris disk excesses of solar-like (F5 - K2/3) 
stars and compared the distribution of metallicities for those with detected disks and those without. 
For their full sample, the Kolmogorov-Smirnov (KS) test indicated a probability of 9\% that the two
samples were drawn from the same parent sample
(although the smaller sample based only on their own metallicity measurements did not 
show such an effect \citep{maldonado15}). The \citet{maldonado12} study was based on 107 disk detections and 145 
non-detections around solar-like stars, mixing warm and cold disks (although most of their disks were detected at 
{\it Spitzer}-MIPS 70 $\micron$). All other studies (excepting \citet{maldonado15}) were based on less than 40 
sources, so it is not surprising that this hint is not apparent in them.

To test this result, we did an analysis comparing detected with undetected disks for the far infrared/cold 
disks\footnote{The warm disk sample shows similar behavior but at lower weight.} based on our trimmed sample 
including spectral types of F0 - K5 (compared with F5 - K2/3 for \citet{maldonado12}; in addition to the 
slightly broader range in spectral type, our sample has systems removed that are sufficiently old that 
debris disks are unexpected as a product of quiescent evolution). Since we reconciled systematic differences in 
metallicity determinations in the process of assembling all the relevant [Fe/H] measurements, our sample is 
equivalent in this sense to the {\it homogeneous} sample of \citet{maldonado12}. The debris disk discriminator 
``detected'' vs.\ ``nondetected'' does not yield a physically well-defined boundary, since it depends on the 
observational circumstances as well as the character of the disk. We therefore considered thresholds 
in terms of our de-aged reduced disk masses. For each threshold, we counted only the detected disk masses above the 
fiducial value (not the upper limits) for one sample, and counted all detections {\it and} upper limits below the 
threshold for the other sample. The useful range of thresholds runs from $\log(M_{\rm reduced}) = -2$ to 
$\log(M_{\rm reduced}) = -4$; above this range there are too few detections and below it too few detections and 
upper limits for a meaningful study of metallicity effects. 

\begin{figure*}[!t]
\begin{center}
\includegraphics[angle=0,scale=0.45]{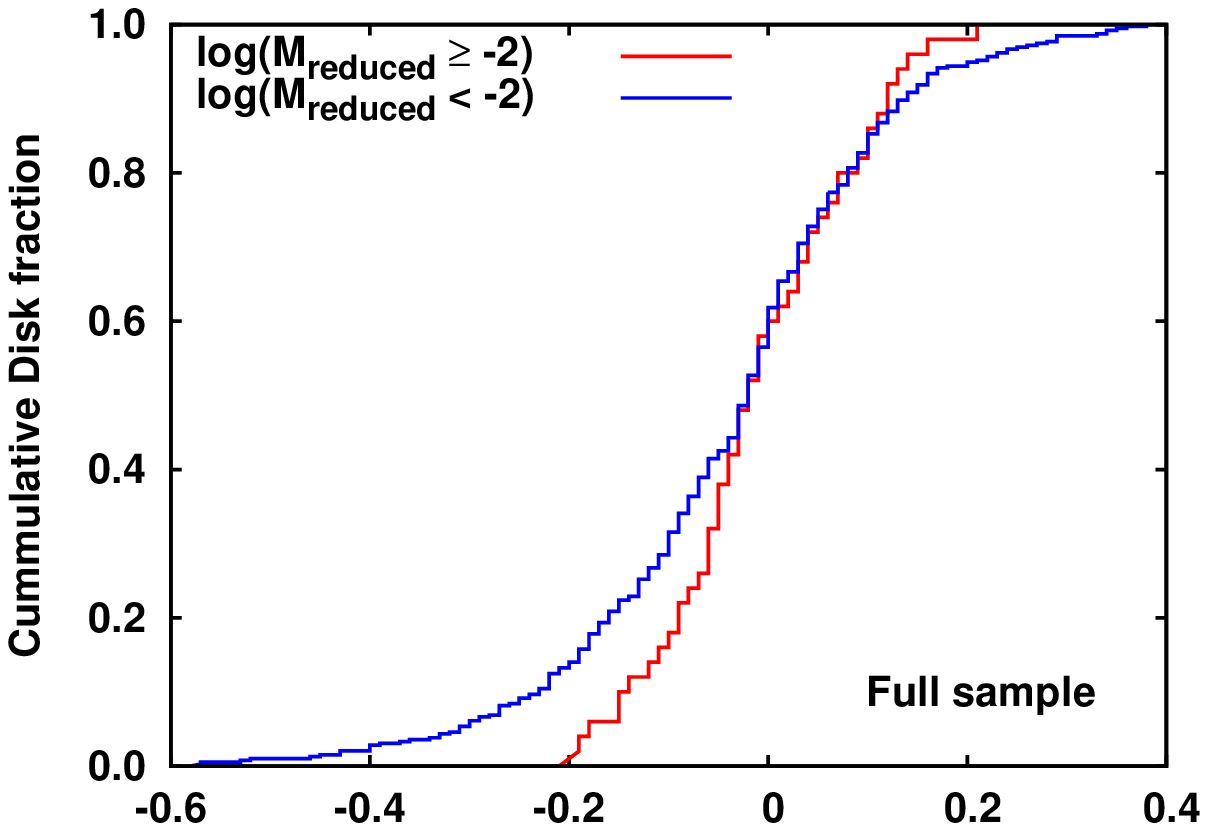}
\includegraphics[angle=0,scale=0.45]{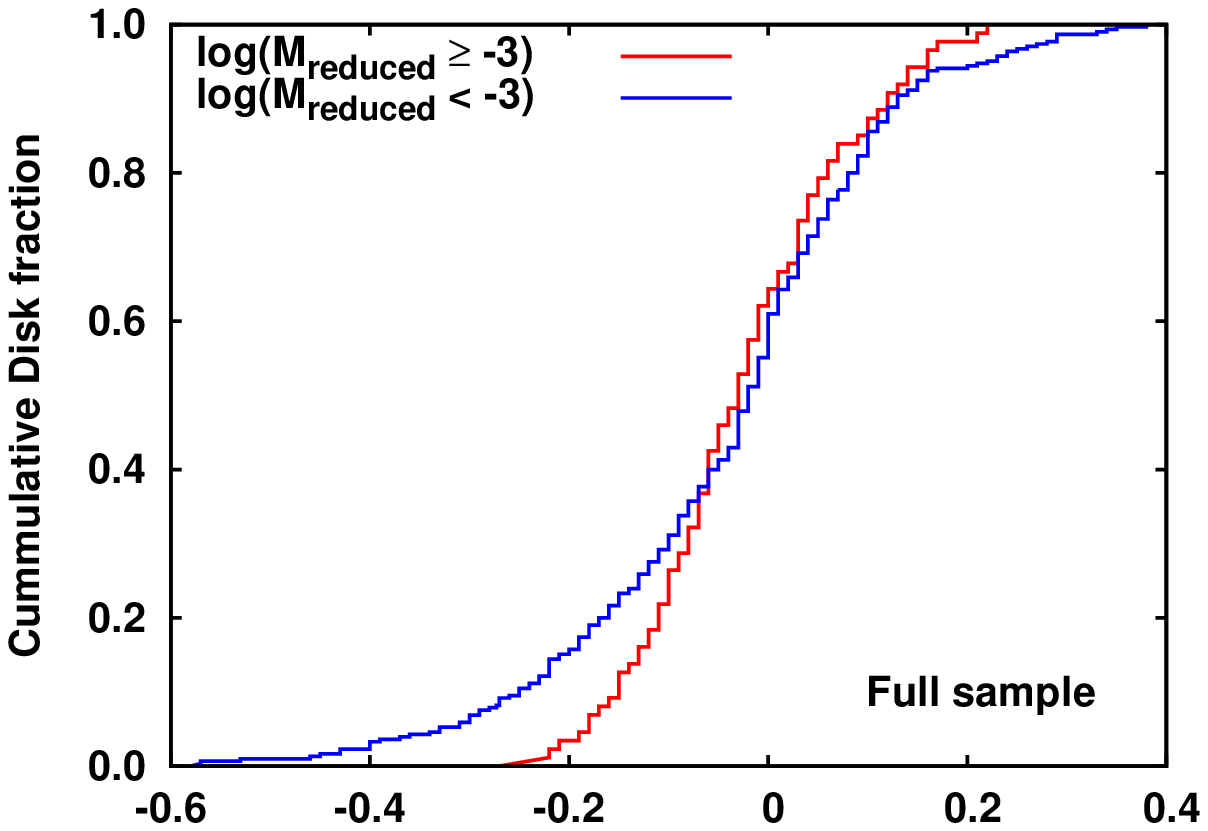}
\includegraphics[angle=0,scale=0.45]{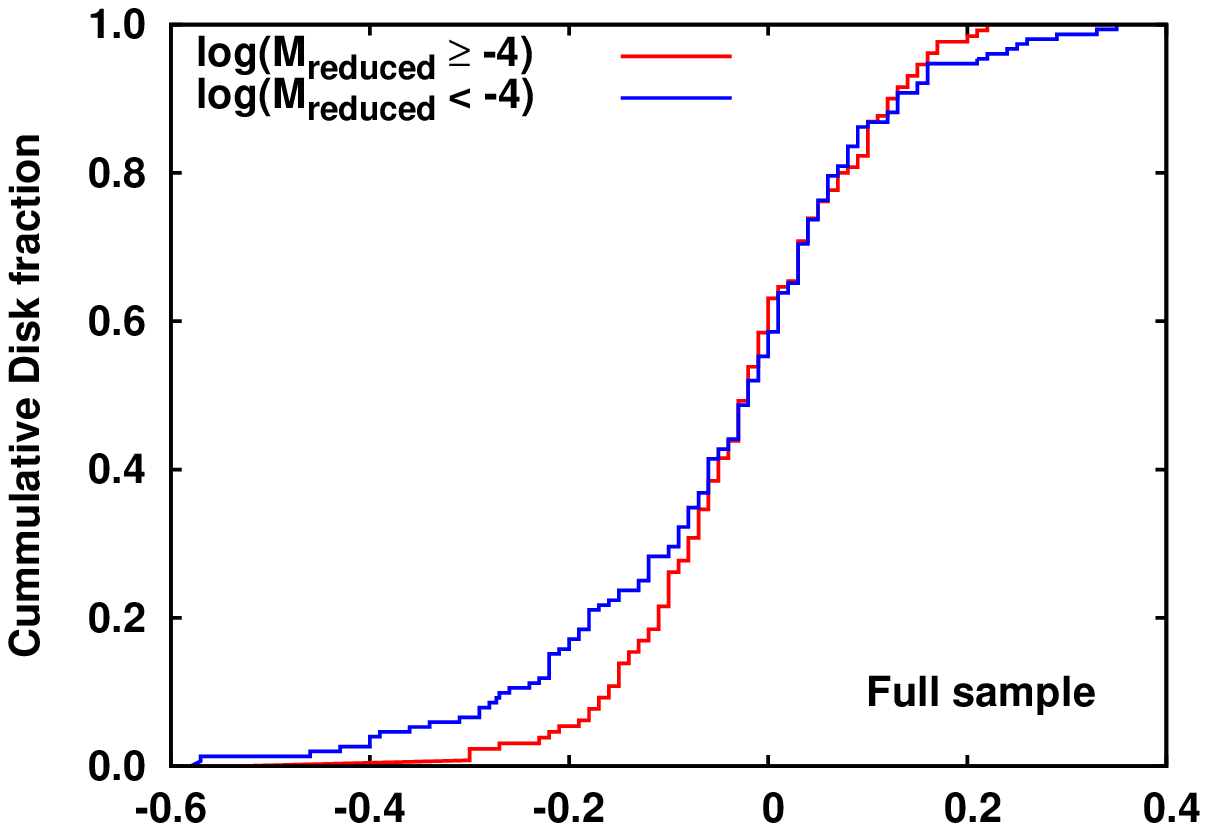}

\includegraphics[angle=0,scale=0.45]{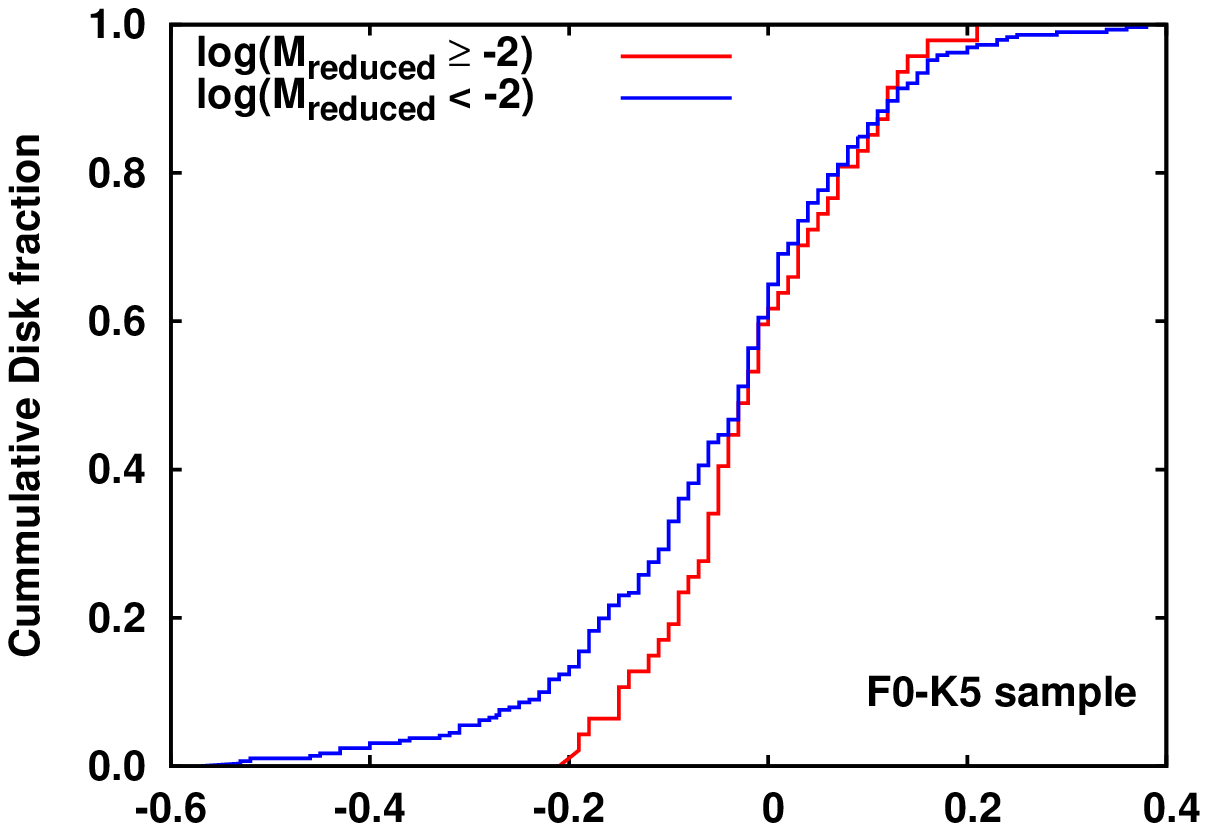}
\includegraphics[angle=0,scale=0.45]{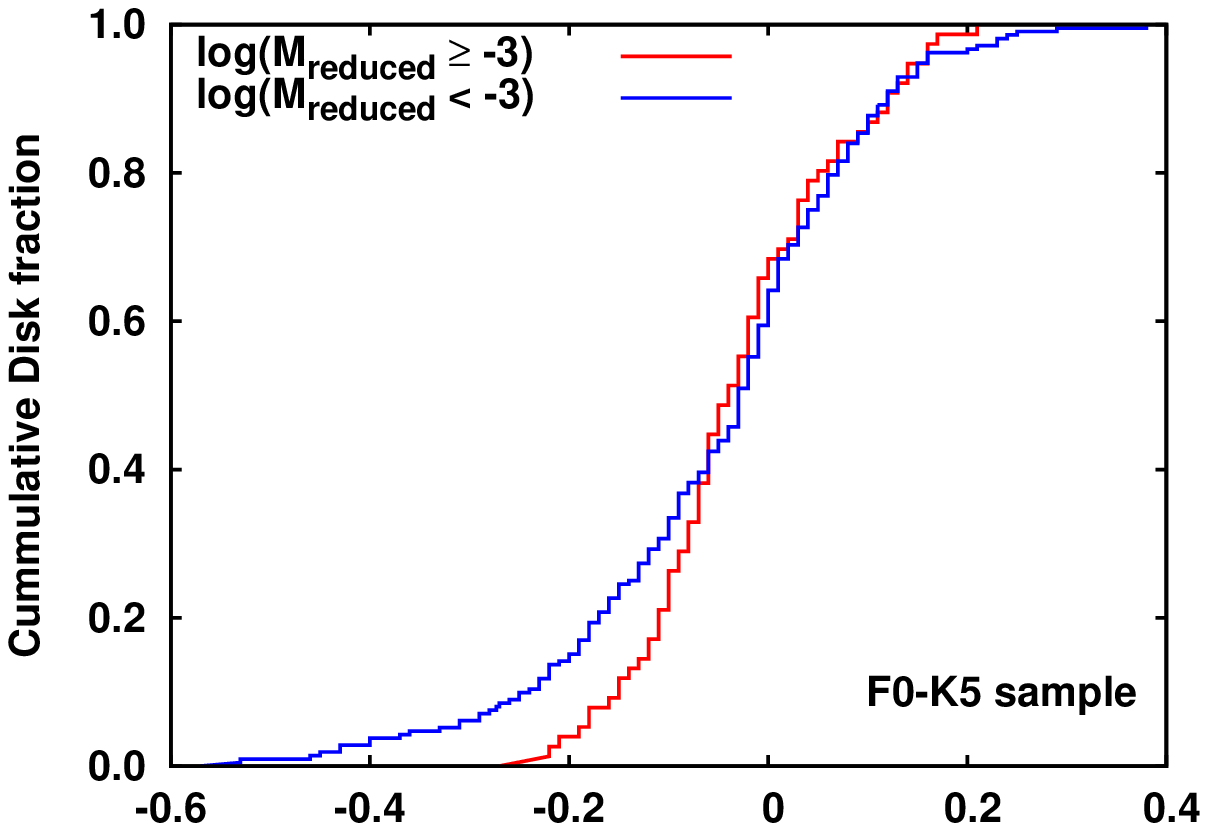}
\includegraphics[angle=0,scale=0.45]{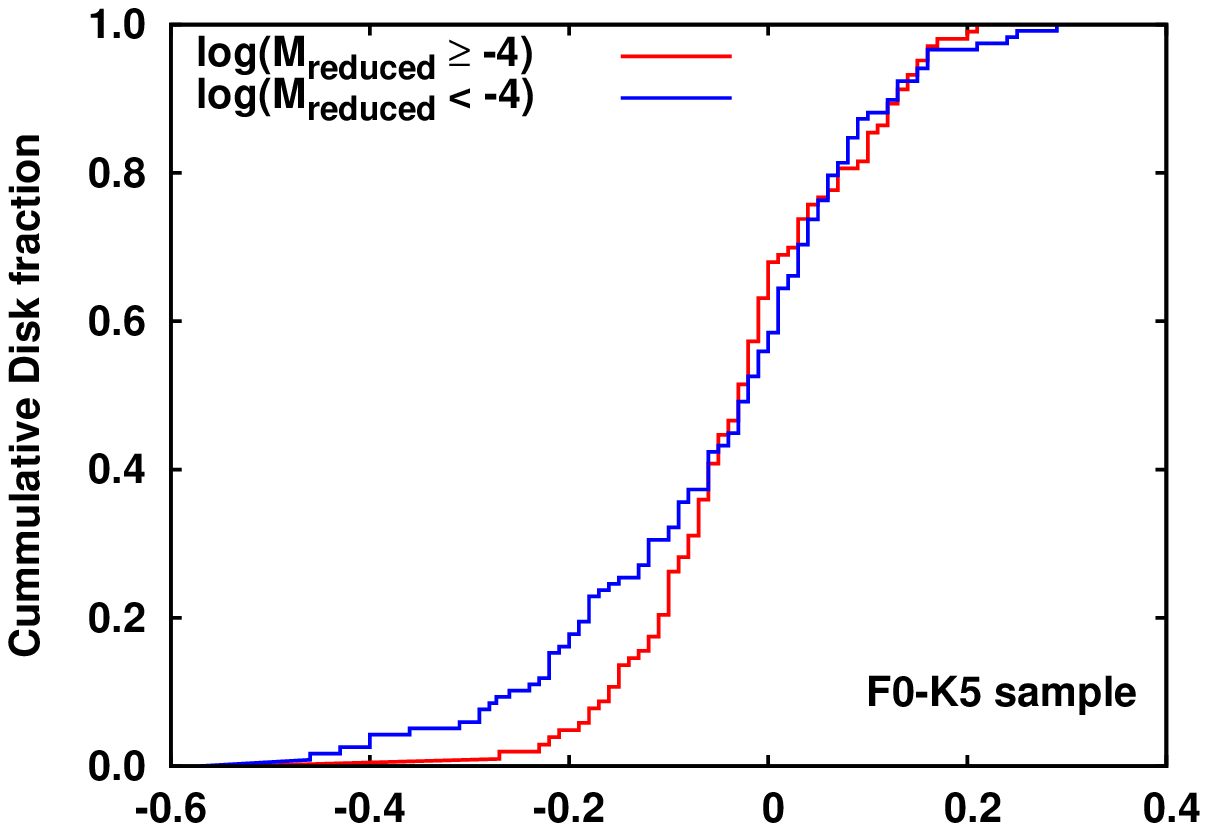}
\caption{The cumulative disk fraction of the cold disk sample at various $M_{\rm reduced}$ 
values for both the full and F0-K5 subsample.}
\label{fig:cCDFs}
\end{center}
\end{figure*}

Figure \ref{fig:cCDFs} shows the results - for completeness, for both the full and trimmed (F0 - K5) cold disk sample. 
For every threshold, the incremental number of systems with increasing [Fe/H] has very similar behavior. 
Above ${\rm [Fe/H]} = -0.1$, the systems below and above the dust mass dividing line have very similar distributions 
indicating that the majority of undetected stars are similar to those with detected disks, but presumably are 
undetected because of less rich planetesimal systems or less vigorous dynamical stirring of those systems 
(the latter probably independent of metallicity). Below ${\rm [Fe/H]} = -0.1$, the distributions of detected and 
undetected systems are dramatically different, with a strong trend toward low metallicity systems not having massive
debris disks. This result therefore confirms and expands the finding by \citet{maldonado12} of a deficit of 
debris-disk-bearing stars over the range $-0.5 \le {\rm [Fe/H]} \le = -0.2$. The K-S test of the trimmed
sample with a cutoff threshold of $\log(M_{\rm reduced}) = -4$ yields $p_{KS}\sim0.15$, in agreement with the \citep{maldonado12} 
result. We also analyzed the samples with the Anderson-Darling (AD) statistic \citep{scholz87}, which gave a $p_{AD}\sim 0.09$. 
The nominal significance of the difference in distributions depends on the way the tested hypothesis is posed: 
1.) if, as above and by \citet{maldonado12}, it is whether the global distributions differ, the result is a strong 
hint of a correlation; however, 2.) if the question is whether there is a tendency against significant debris sytems 
for stars with less than solar [Fe/H], the statistical significance of the difference (e.g., comparing distributions 
for ${\rm [Fe/H]} < 0$) is greater.

\subsection{Metallicity effects in our warm and cold samples}

\begin{figure*}[!t]
\begin{center}
\includegraphics[angle=0,scale=0.45]{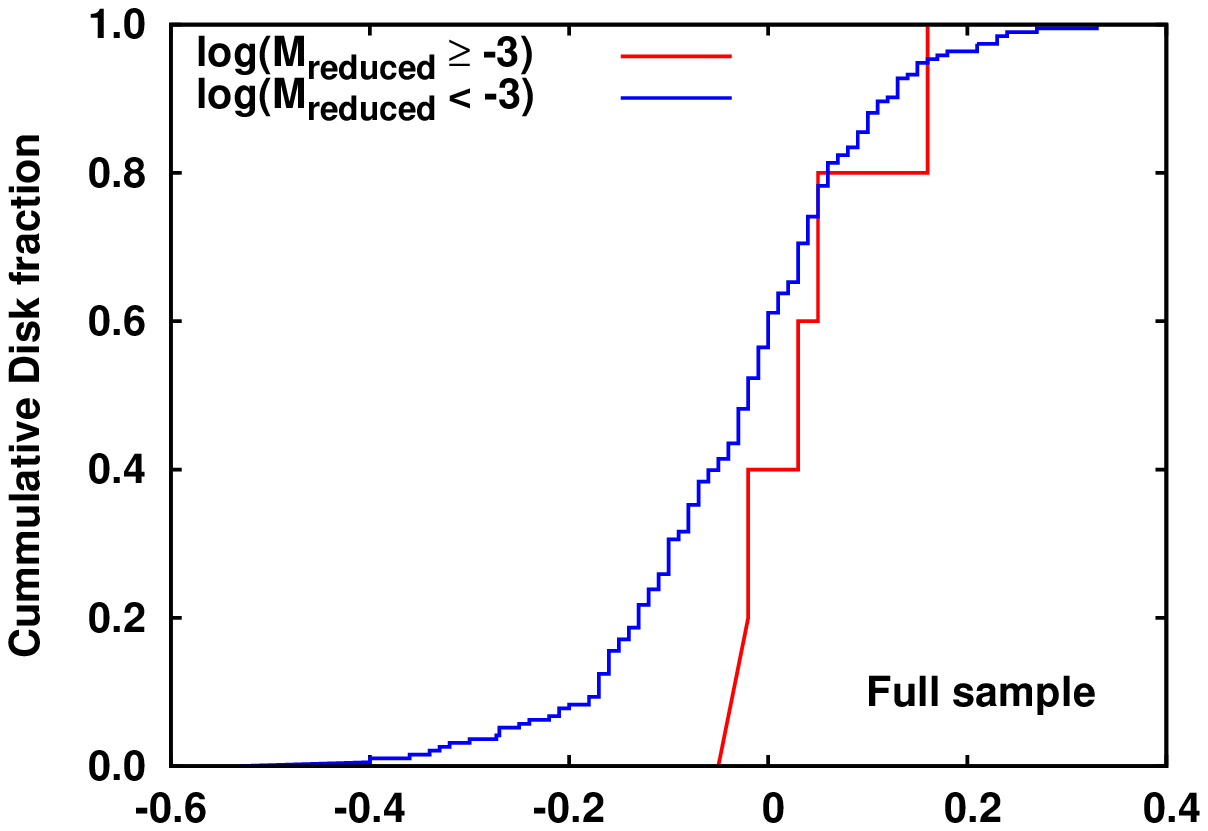}
\includegraphics[angle=0,scale=0.45]{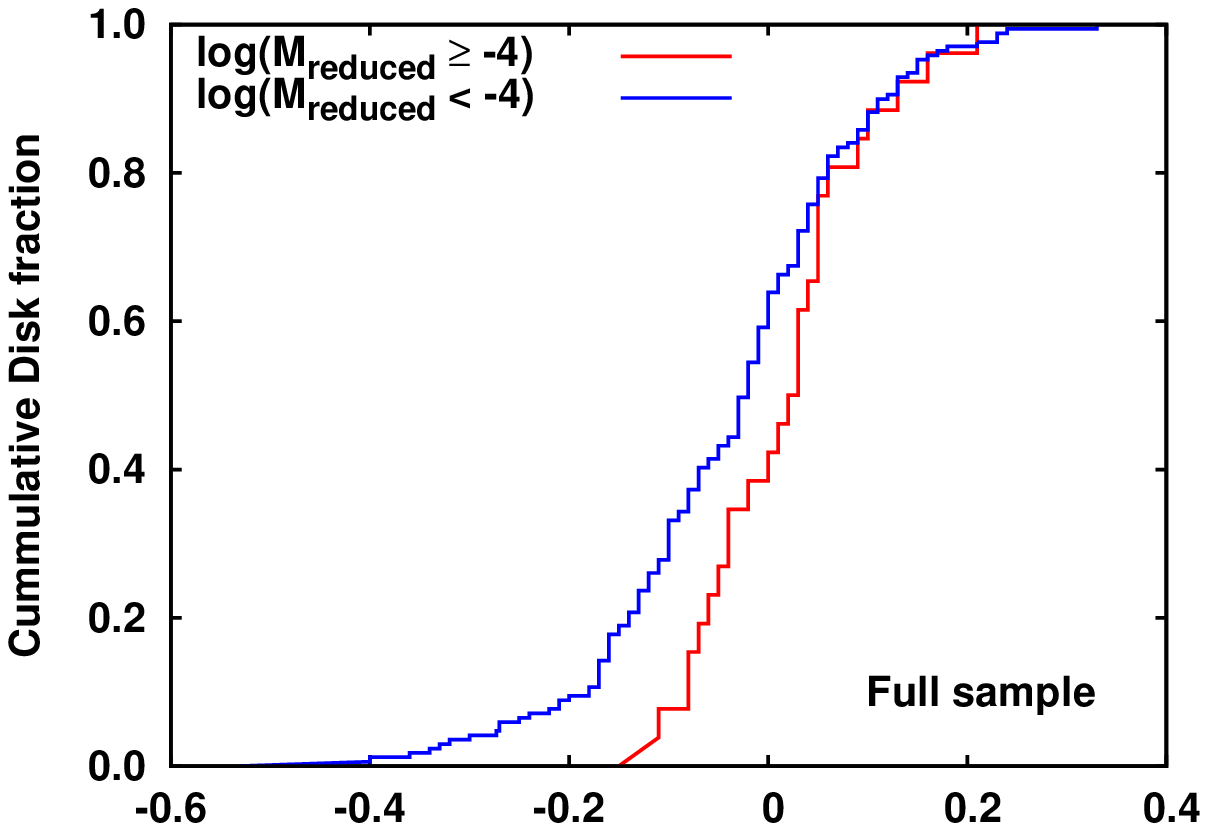}
\includegraphics[angle=0,scale=0.45]{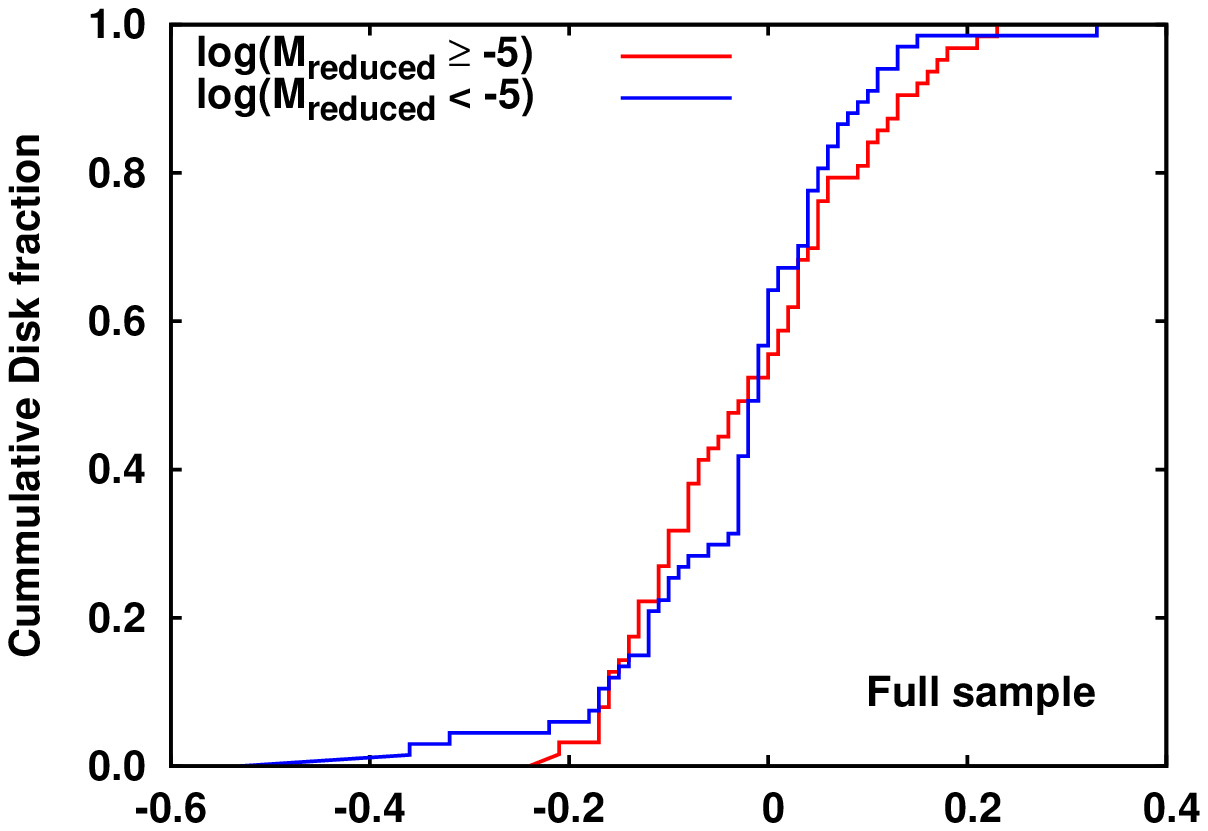}

\includegraphics[angle=0,scale=0.45]{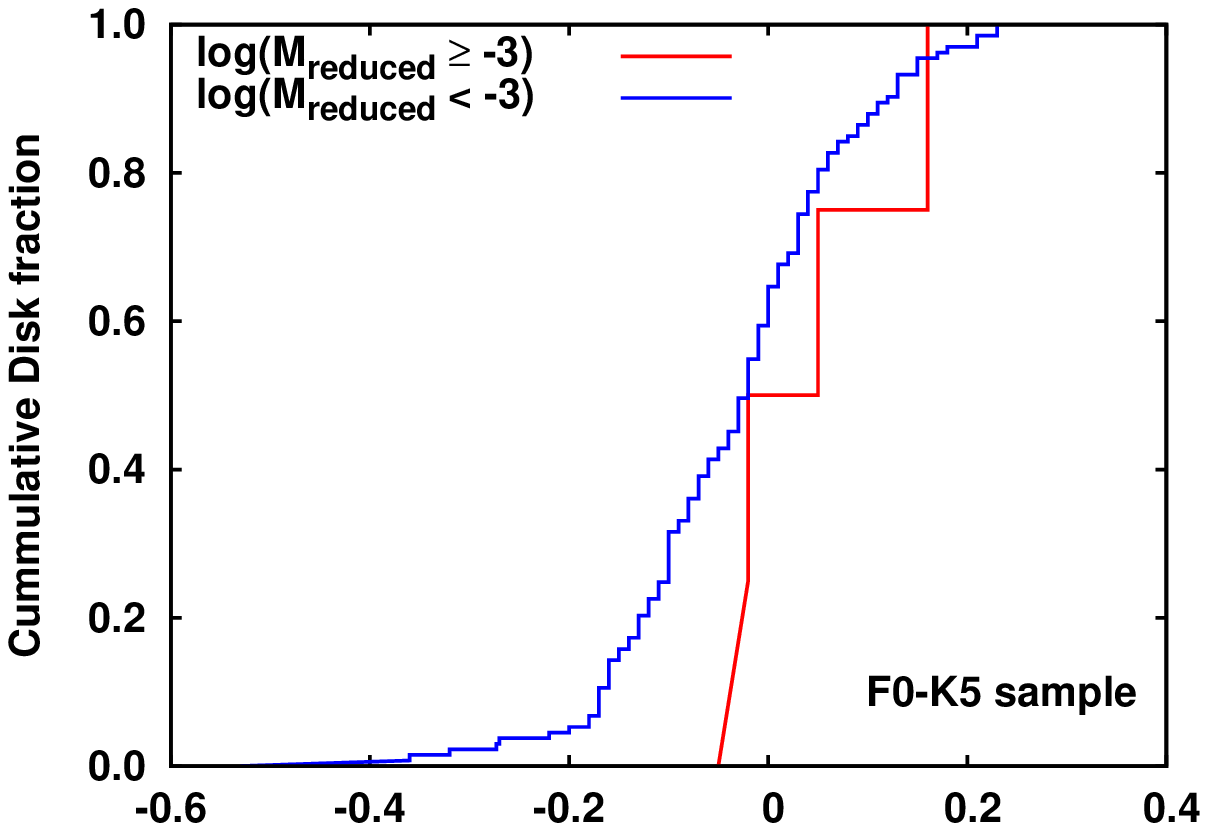}
\includegraphics[angle=0,scale=0.45]{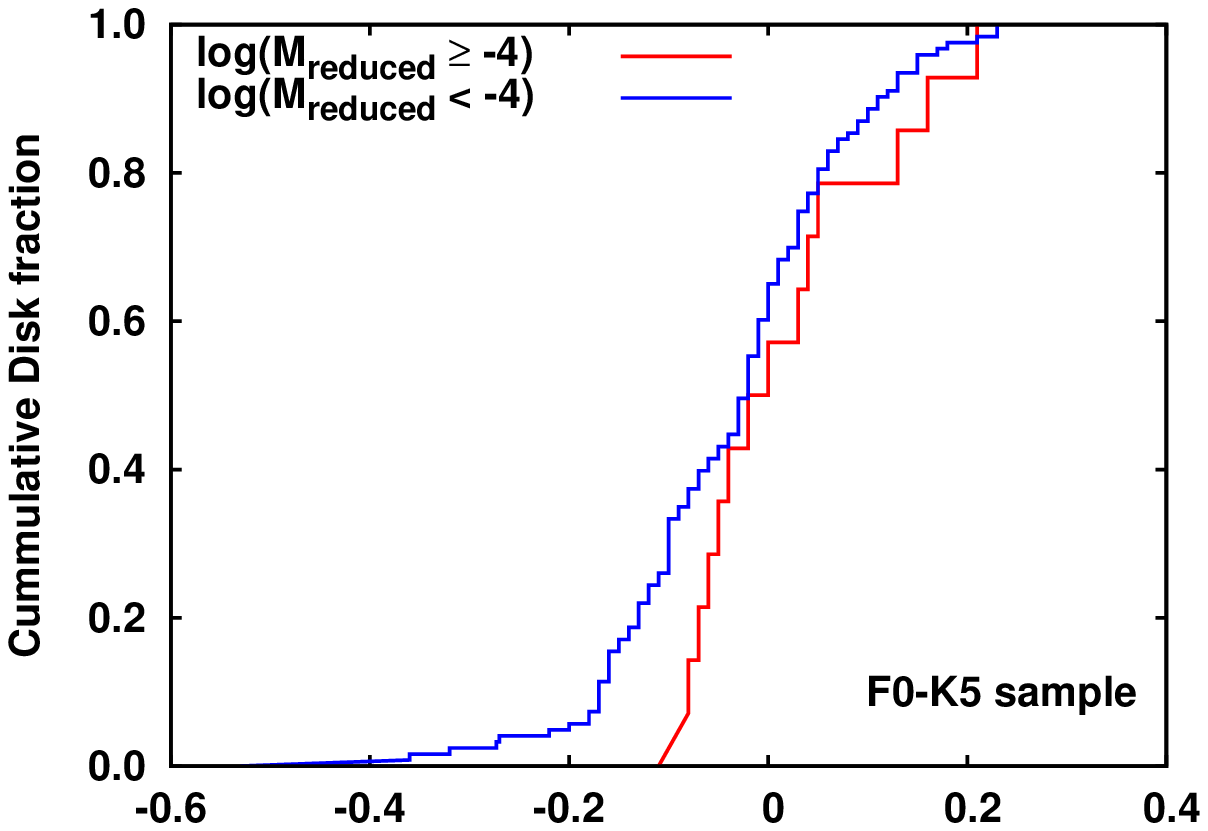}
\includegraphics[angle=0,scale=0.45]{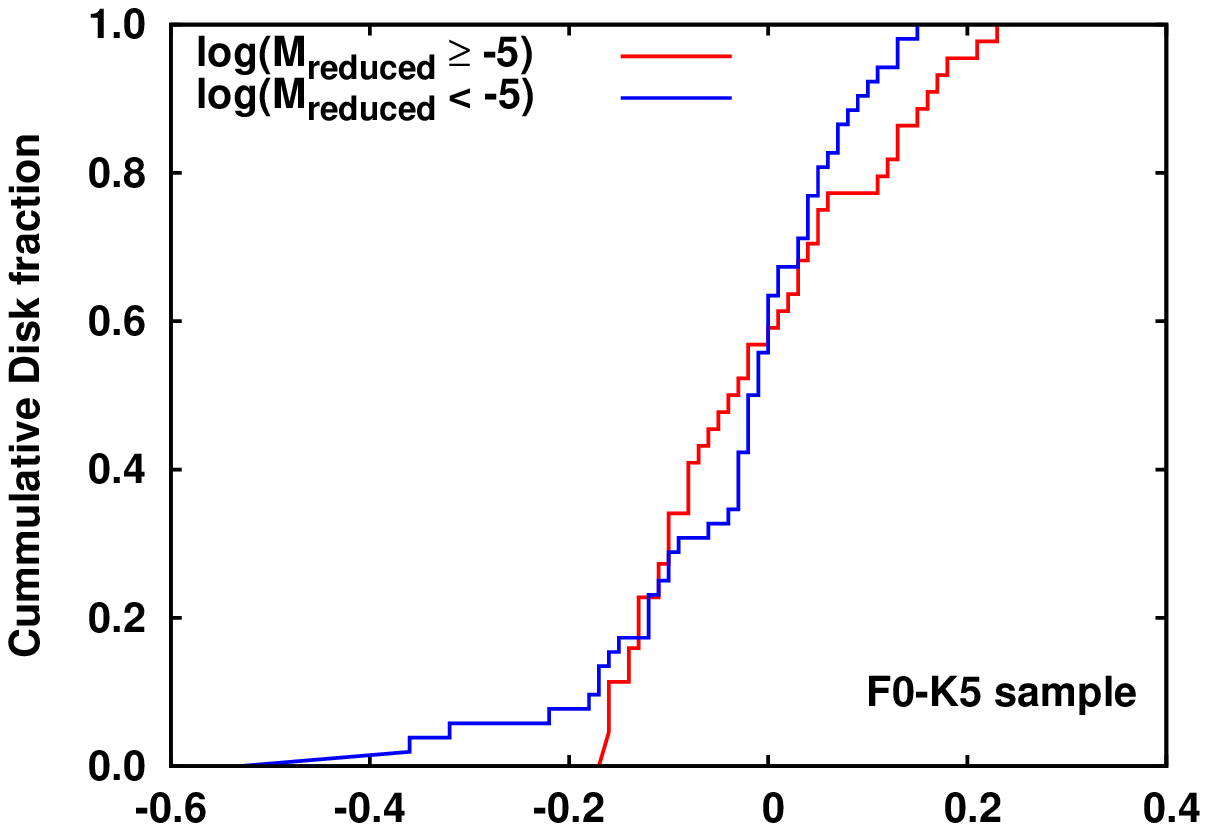}
\caption{The cumulative disk fraction of the warm disk sample at various $M_{\rm reduced}$ 
values for both the full and F0-K5 subsample.}
\label{fig:wCDFs}
\end{center}
\end{figure*}

\begin{figure*}[!t]
\begin{center}
\includegraphics[angle=0,scale=0.45]{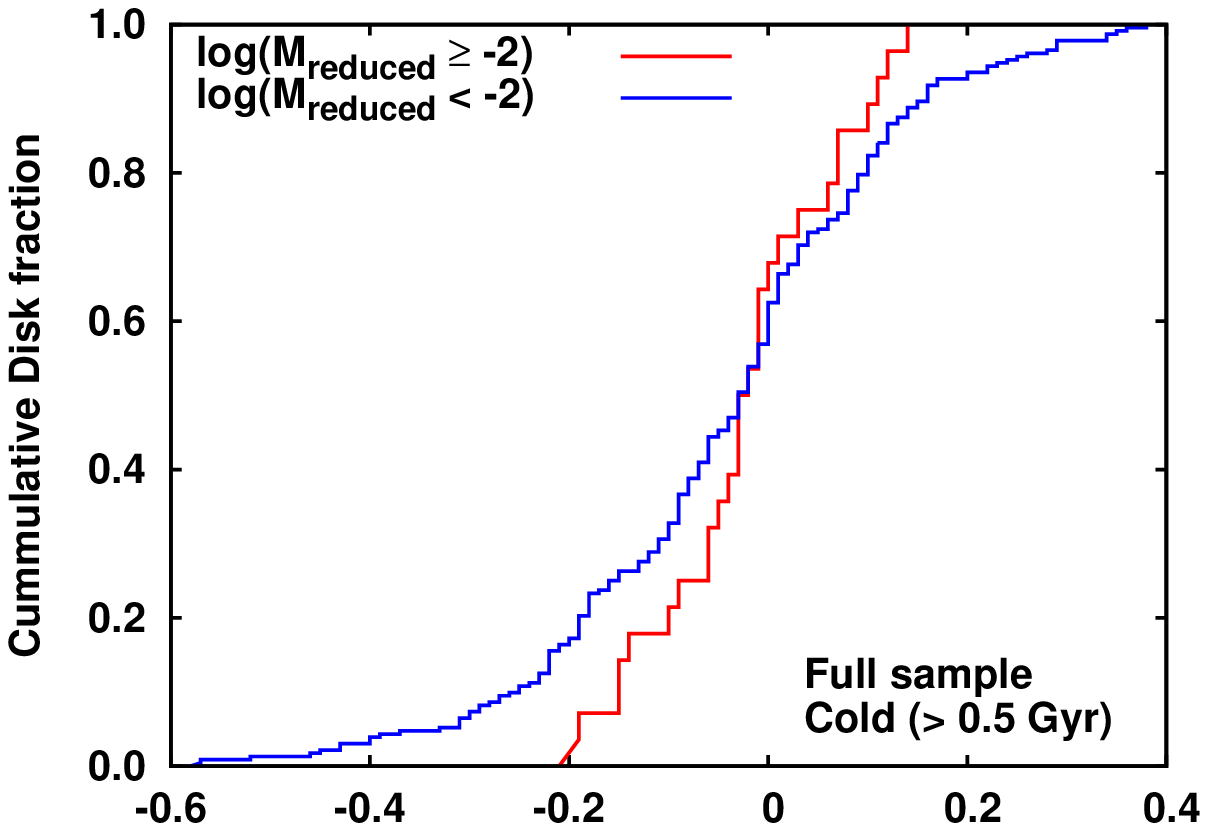}
\includegraphics[angle=0,scale=0.45]{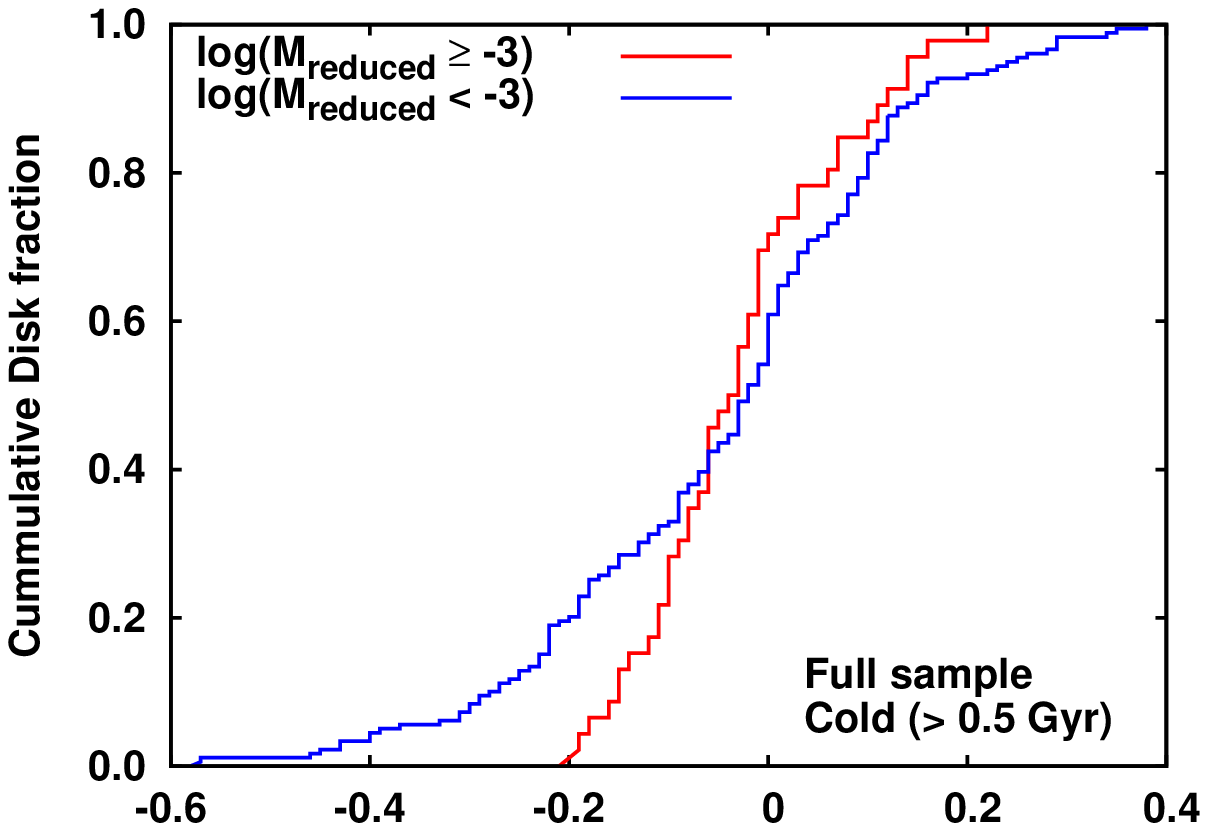}
\includegraphics[angle=0,scale=0.45]{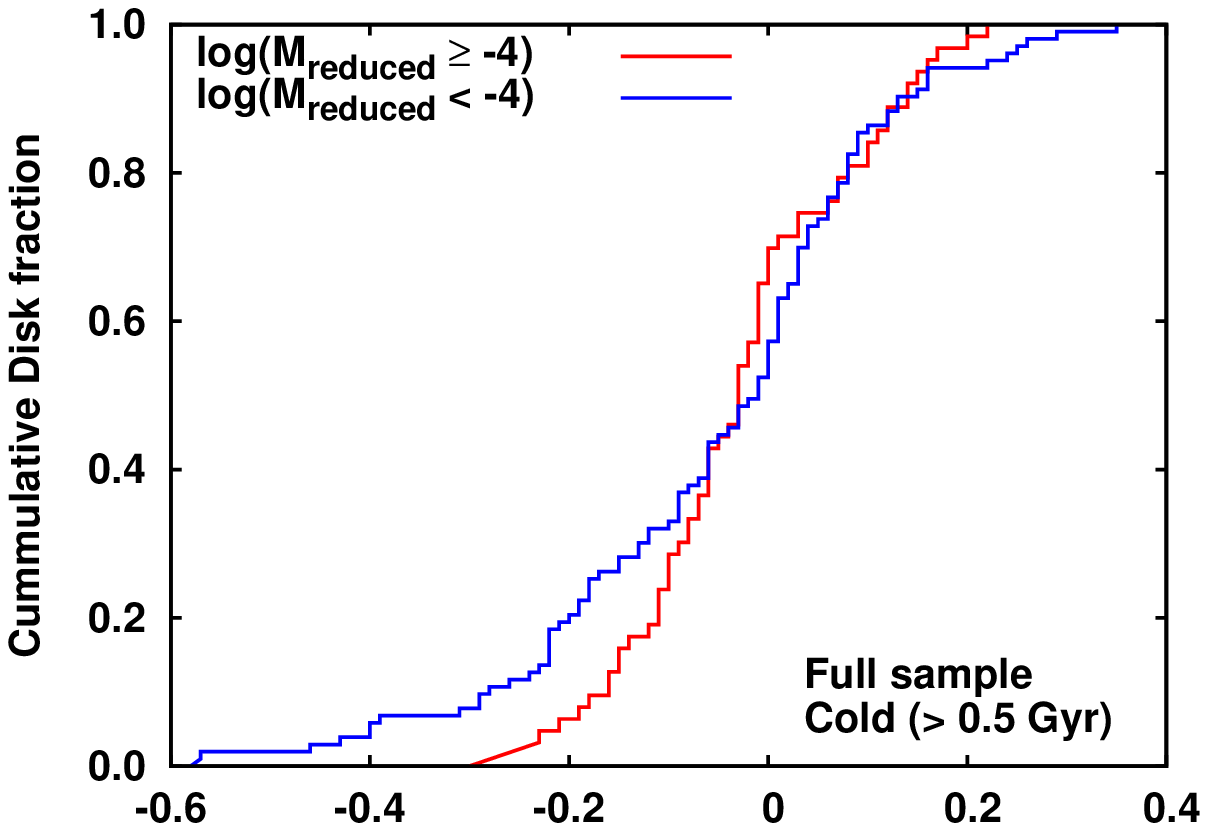}

\includegraphics[angle=0,scale=0.45]{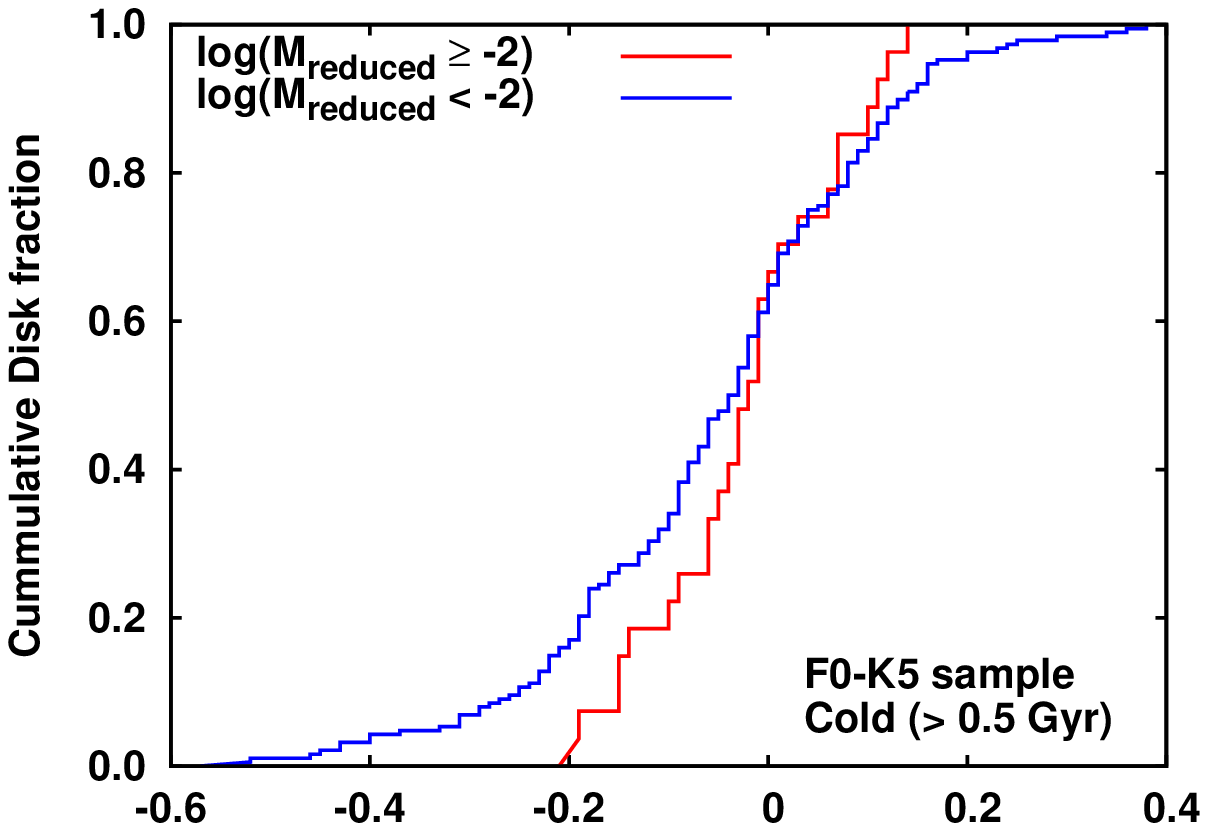}
\includegraphics[angle=0,scale=0.45]{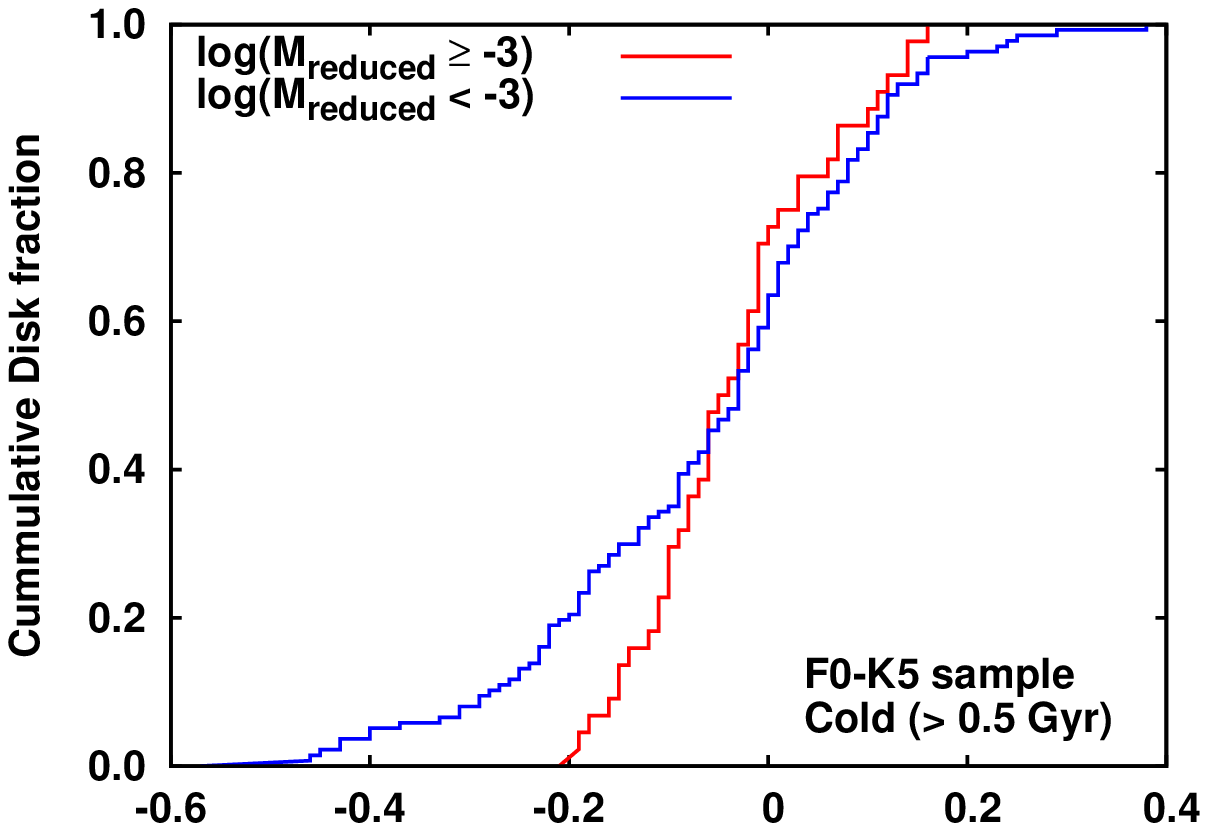}
\includegraphics[angle=0,scale=0.45]{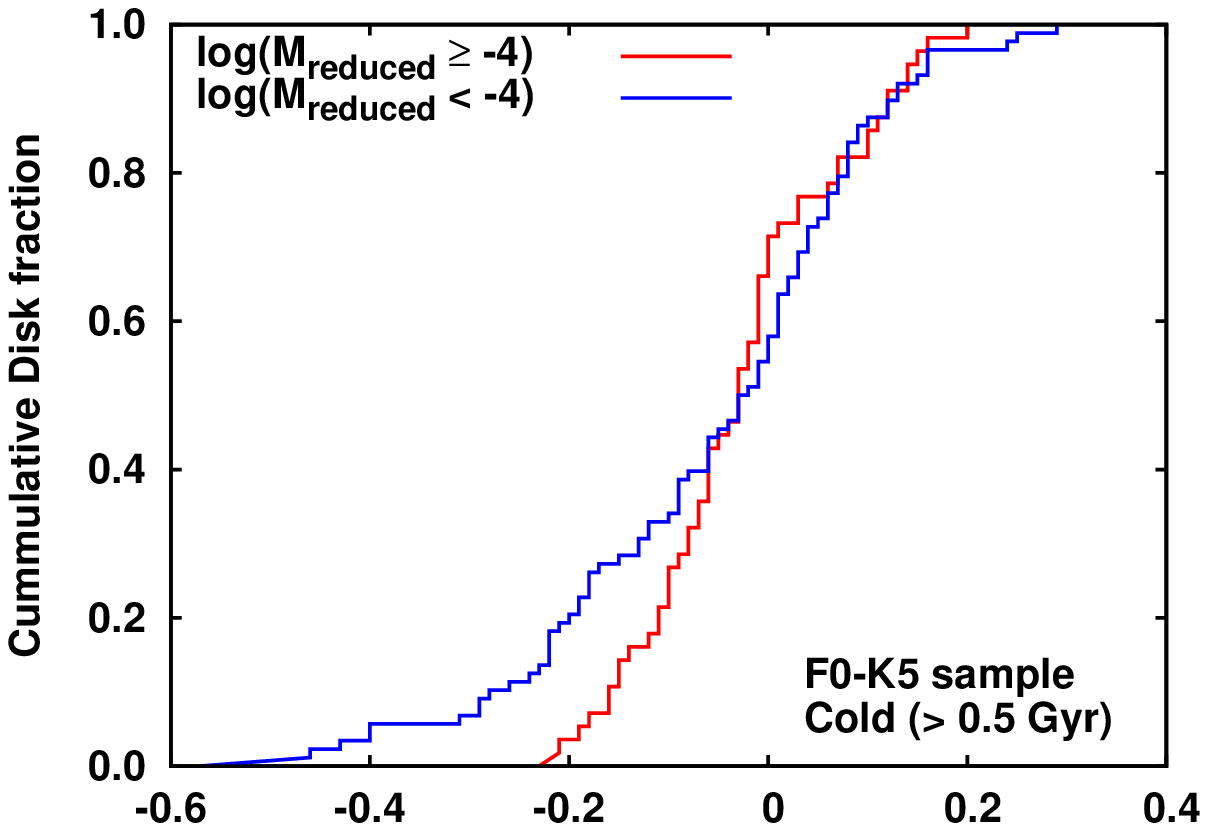}
\caption{The cumulative disk fraction of the cold disk sample at various $M_{\rm reduced}$ 
values for both the full and F0-K5 subsample, including only sources that are older than 0.5 Gyr.}
\label{fig:cCDFsold}
\end{center}
\end{figure*}

In Figure \ref{fig:wCDFs}, we show the same distribution functions for the warm
disk sample. At a reduced mass cut of $\log(M_{\rm reduced}) = -4$ a similar discrepancy can be seen between the 
distributions as for the cold disk systems, with $p_{KS}\sim0.07$. The AD statistic gave a higher $p_{AD}\sim0.2$ value
for the warm sample, likely do to the low number of sources (14) in one of the samples. Finally, we repeated this test 
for cold disks around stars older than 0.5 Gyr. Since we required an age below this value for the warm disks, this is a 
completely independent sample. The results, exhibited in Figure \ref{fig:cCDFsold}, again demonstrate, with $p_{KS}\sim0.11$
($p_{AD}\sim0.08$), an absence of disks with ${\rm [Fe/H]} \le -0.2$. The very similar behavior in {\it both} Figures 
\ref{fig:wCDFs} and \ref{fig:cCDFsold} supports that the lack of disks at low metallicity is a significant trend. 

\subsection{Survival Analysis}

The previous section found evidence for an effect in terms of the metallicity as a function of the disk properties. 
We now probe in the orthogonal sense, i.e., the disk properties as a function of the metallicity. We divide 
each of the warm and cold samples into three equal-sized bins in metallicity. For the warm sample, 
the demarcations are at ${\rm [Fe/H]} = -0.085$ and $0.034$, with 66 disks in each bin, including the 
censored data (upper limits). This roughly defined a metal poor, a solar metallicity, and a metal rich 
group. For the cold disk sample, the demarcations were at almost the same values, ${\rm [Fe/H]} = -0.09$ 
and $0.028$, with 154 disks in each group, including censored data (upper limits). We will investigate the 
mass distribution of the disks in each of these bins.

The Kaplan-Meier (K-M) survival estimator \citep{km58} provides a systematic method for doing so. Originally developed 
for biological studies (hence its slightly inappropriate name), survival analysis has been adopted for 
astronomical data analysis \citep[e.g.,][]{feigelson85}, where it is useful for randomly picked datasets. 
We used the statistical routine {\tt survfit} in the {\tt NADA} package of {\tt R} to calculate 
the K-M survival estimate with increasing dust mass within each of these groups for the systems at their 
current ages and the de-aged 1 Myr old total dust mass distributions. The routine also calculates the 
95\% confidence intervals of the survival estimates.

\begin{figure*}[t]
\begin{center}
\includegraphics[angle=0,scale=0.55]{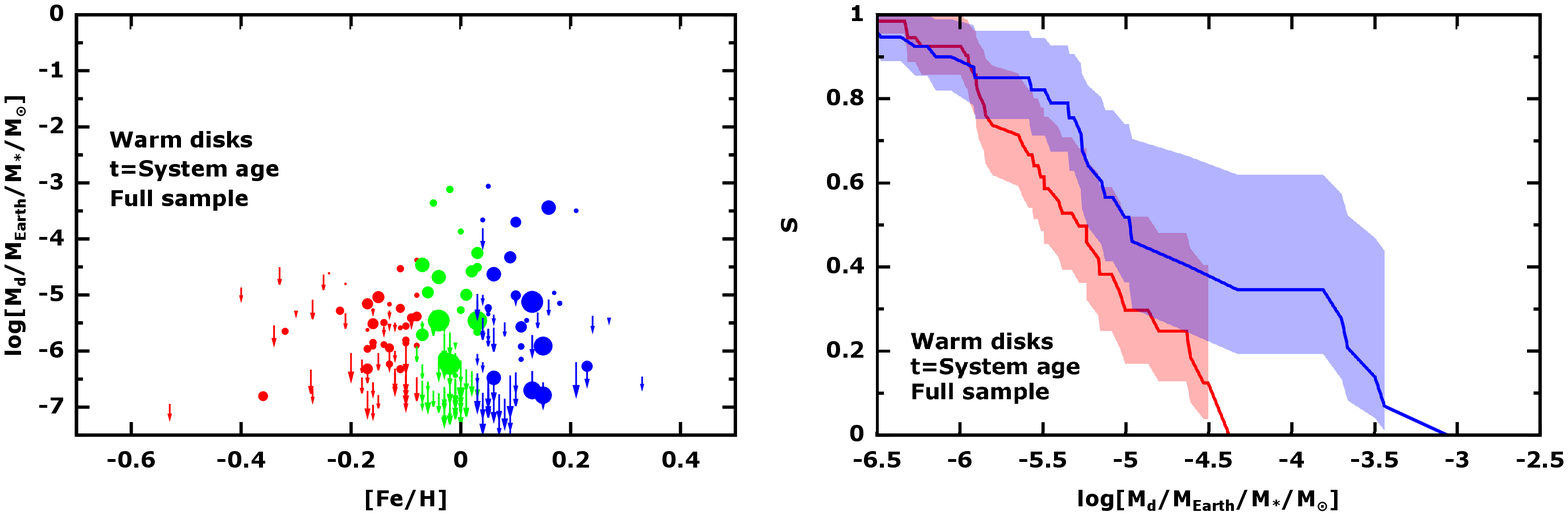} 
\includegraphics[angle=0,scale=0.55]{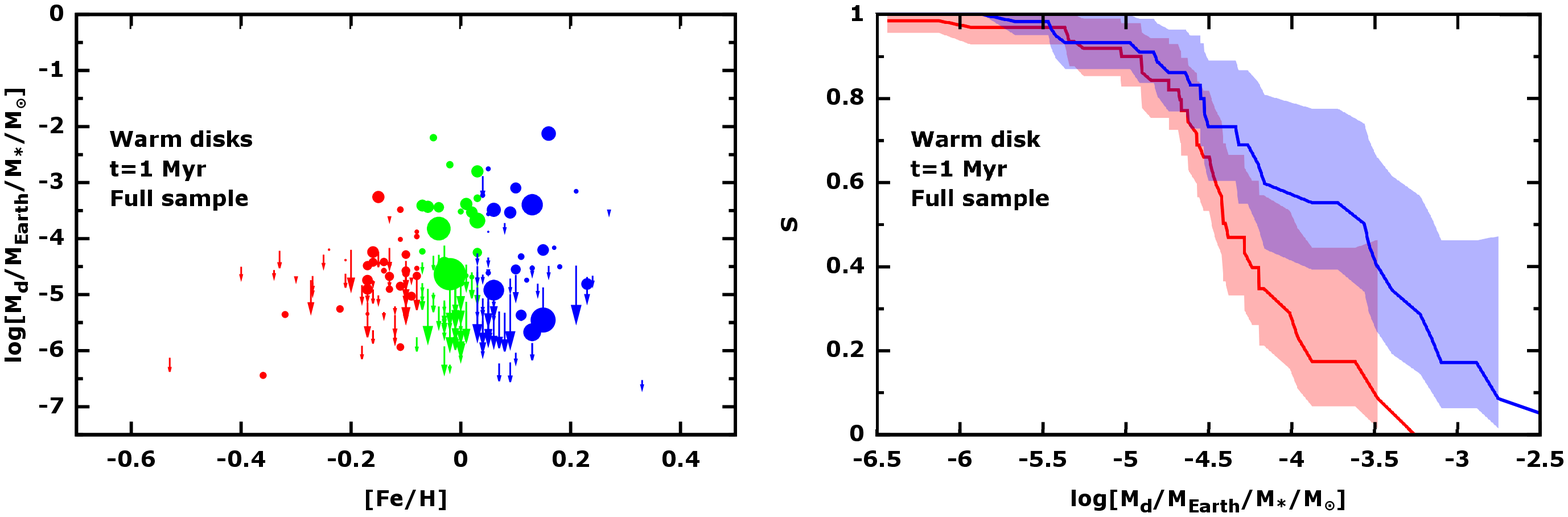}

\includegraphics[angle=0,scale=0.55]{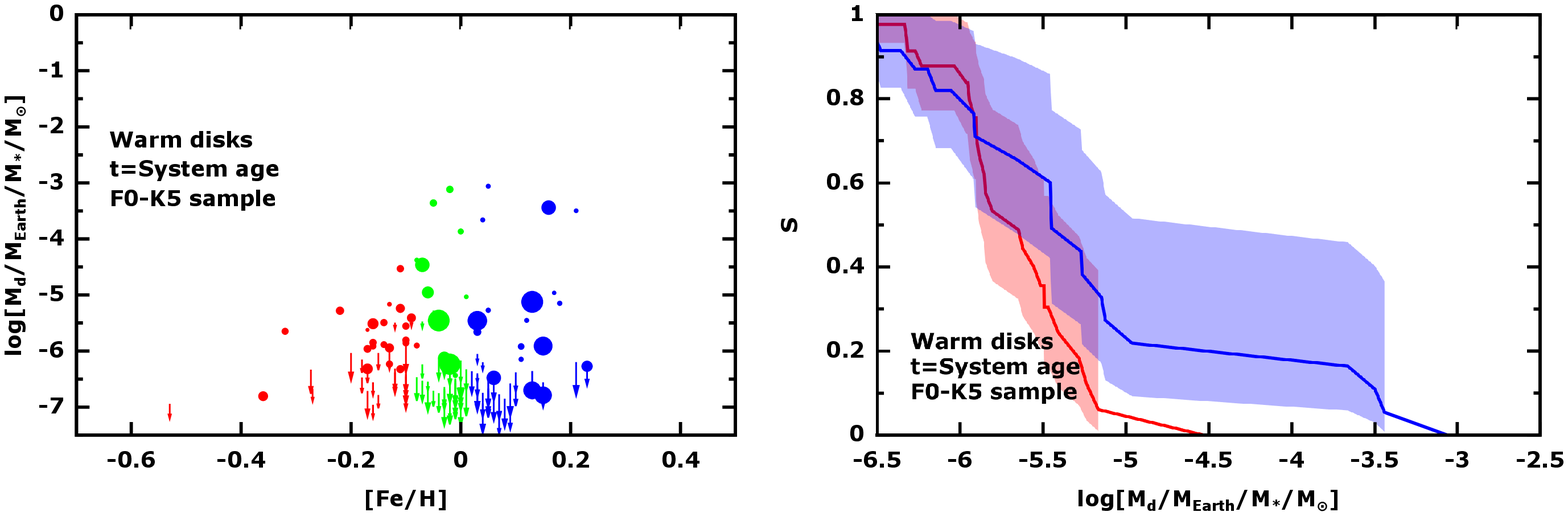} 
\includegraphics[angle=0,scale=0.55]{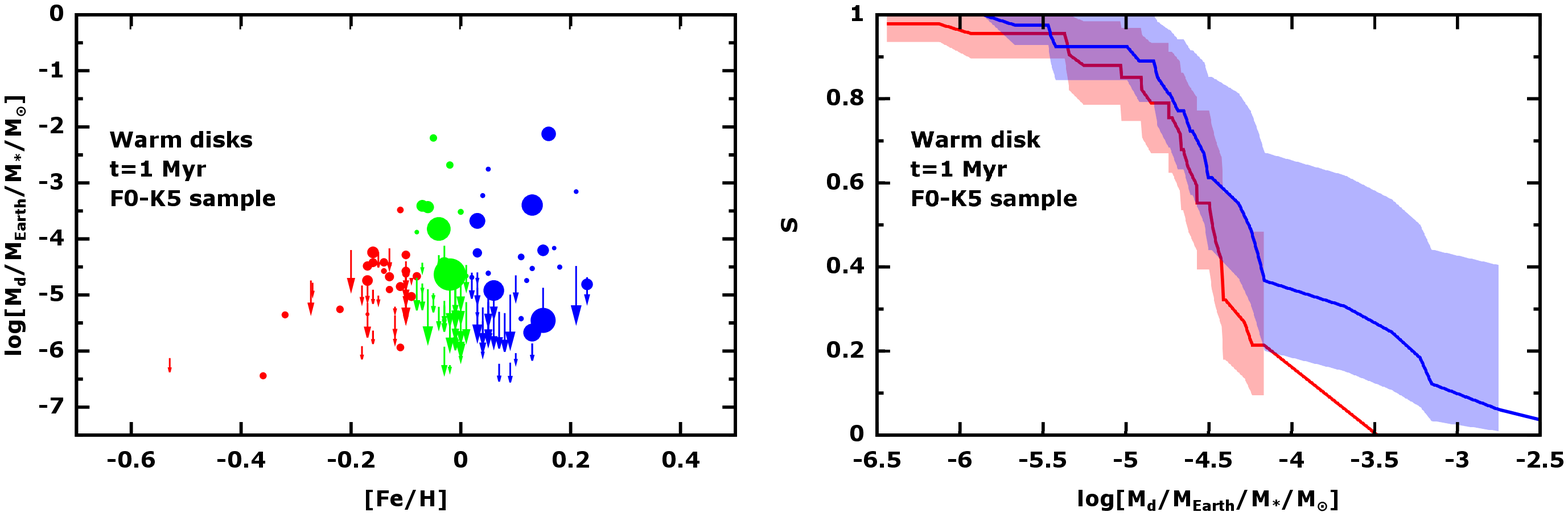}
\caption{Survival analysis of the disk dust mass distributions for the warm debris disks, with the samples divided into 
three metallicity bins with equal numbers of members. The first two rows show plots for the full sample, while the last 
two rows show plots for the F0-K5 subsample. In the left plots, the sizes of the points are inversely in proportion to 
their errors - the largest points have the smallest errors - to guide the eye in weighting the points. The error bands in the KM analyses to the right give 
a 95\% confidence interval at each disk dust mass. For clarity, we only show the KM functions of the low and high
metallicity bins; the KM function of the solar-metallicity bin roughly agrees with the KM function of the higher 
metallicity one. At 1 Myr, the metal-poor bin has a significantly steeper survival function, with lower 
mass disks than the solar and metal-rich groups. This difference shows that the metal poor bin is drawn from a 
different parent disk dust mass distribution than the metal rich bin. The difference grows with increasing dust mass, 
consistent with the presence of a metallicity-dependent upper envelope to the dust mass.}
\label{fig:warmkm}
\end{center}
\end{figure*}

\begin{figure*}[t]
\begin{center}
\includegraphics[angle=0,scale=0.55]{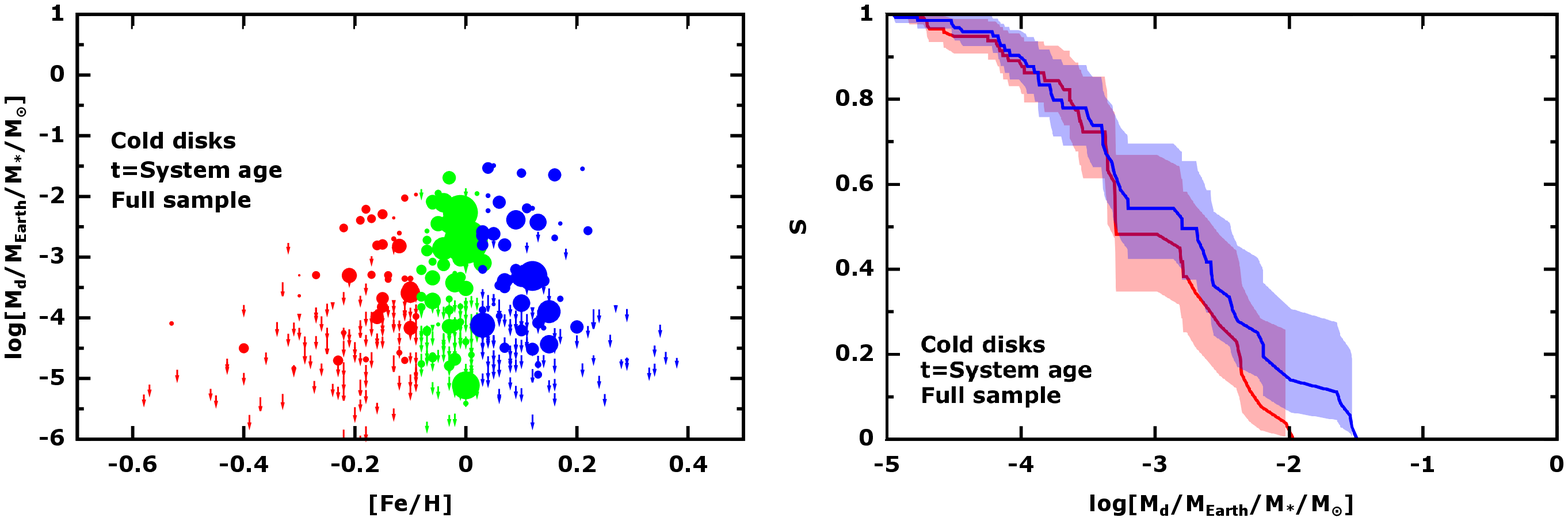}
\includegraphics[angle=0,scale=0.55]{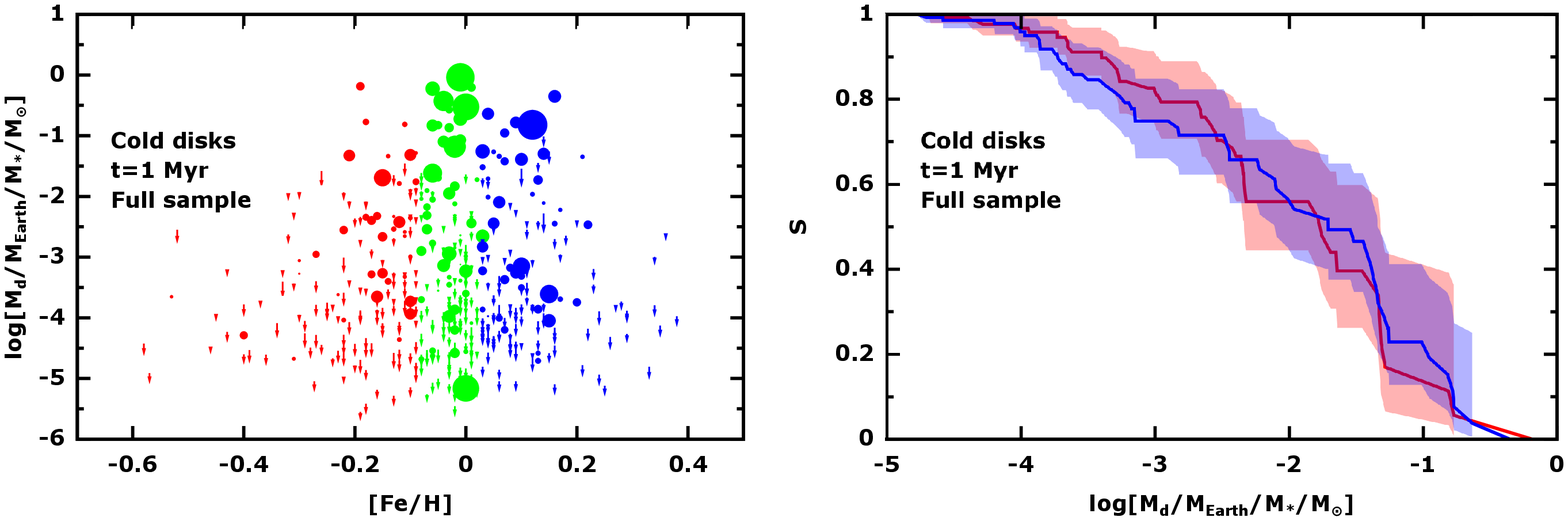}

\includegraphics[angle=0,scale=0.55]{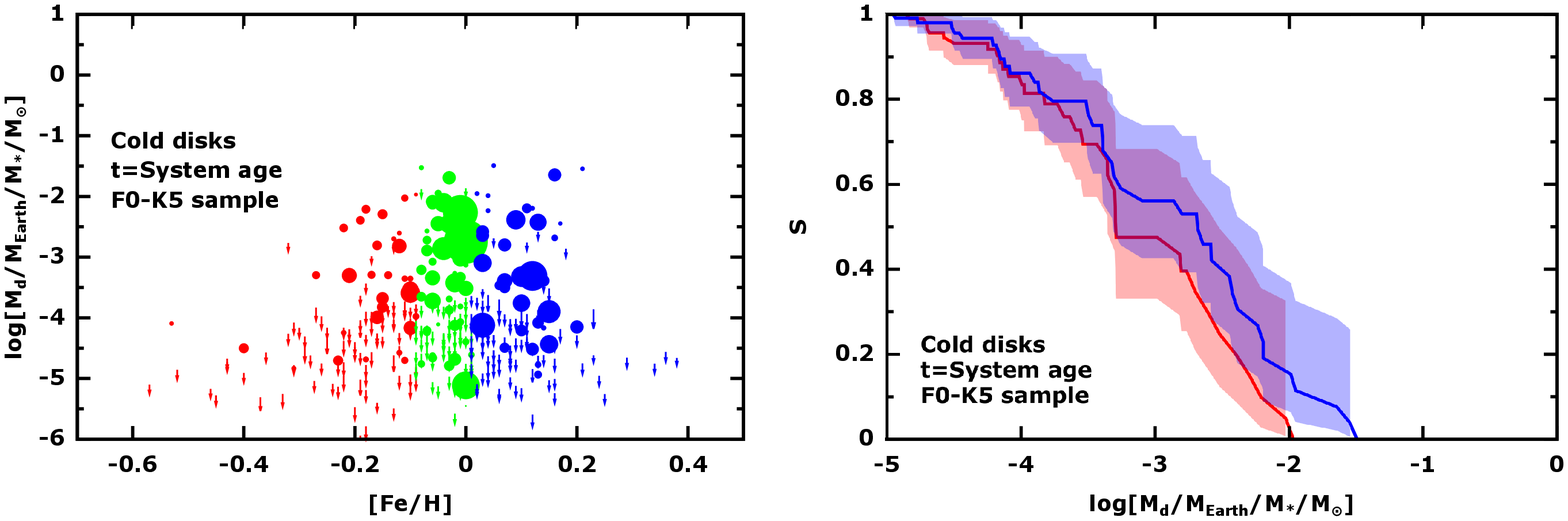}
\includegraphics[angle=0,scale=0.55]{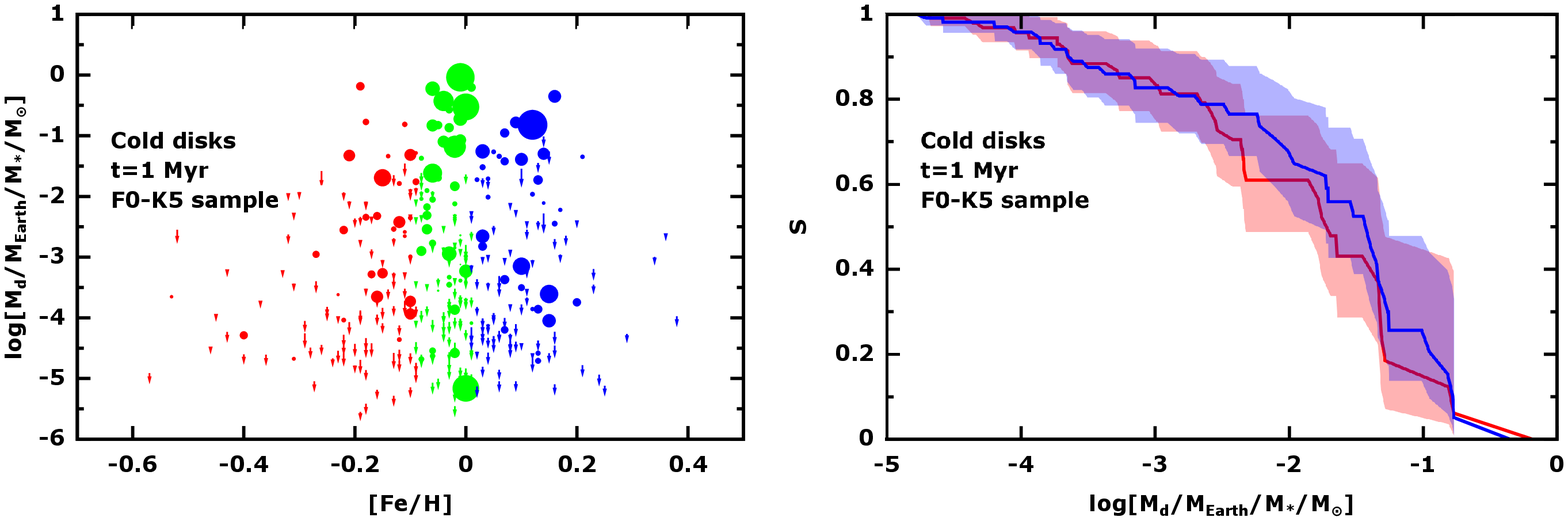}
\caption{Survival analysis of the disk dust mass distributions for the cold debris disks, with the samples divided 
into three metallicity bins with equal numbers of members. The first two rows show plots for the full sample, 
while the last two rows show plots for the F0-K5 subsample. In the left plots, the sizes of the points are inversely 
in proportion to their errors - the largest points have the smallest errors - to guide the eye in weighting the points. The error bands in the KM analyses to the 
right give a 95\% confidence interval at each disk dust mass. For clarity, we only show the KM 
functions of the low and high metallicity bins; the KM function of the solar-metallicity bin roughly agrees with 
the KM function of the higher metallicity one.}
\label{fig:coldkm}
\end{center}
\end{figure*}

We show the disk dust mass vs.\ metallicity distribution of the lower and higher metallicity bins for the warm disk 
sample and the survival functions of the sources in mass space at the current ages and for the de-aged results 
in Figure \ref{fig:warmkm} (the intermediate metallicity bin generally falls near the high metallicity case). 
The data point sizes in the distribution plots are proportional to the inverse of their metallicity errors
to guide the eye in weighting each point.
The top row of the figure shows the plots for the full, while the bottom is for the trimmed F0-K5 spectral-type sample. 
There is a hint of a difference with metallicity prior to de-aging the systems. Applied to the de-aged results, this 
test shows that at 1 Myr, near the time of their formation, there is perhaps a modest trend with metallicity for the 
least massive disks, but a significant trend develops with increasing mass. It is especially noteworthy that the highest 
mass system in the metal poor sample has a reduced mass of $\log(M_{\rm reduced}) = -3.5$, while there are 4 sources 
more massive than this value in each of the higher metallicity bins. This behavior suggests a metallicity-dependent 
upper limit to the disk masses, which seems plausible from visible inspection of the left plot. 

We show the disk dust mass vs.\ metallicity distribution of the three metallicity bins in the cold sample at both ages 
and the survival functions of the sources in dust mass space in Figure \ref{fig:coldkm}. The survival functions agree 
with each other within their 95\% confidence intervals for all the bins, although that for the low metallicity bin 
tends to fall below that for the high metallicity one for the higher-mass disks. There may be a trend at the 
$\sim 1\sigma$  level for low metallicity systems not to have massive disks, but this result is by itself not 
statistically significant.

\subsection{Form of the Disk Mass vs.\ [Fe/H] Relation}

We have tried to extract more information about the relationship between disk mass and stellar metallicity by carrying out 
linear regression on the low and high metallicity portions of our sample. There are significant challenges in doing so: the 
dataset we are analyzing has errors in both variables (metallicity and system dust mass) and also has censoring (upper 
limits on the dust mass) for some of the data. The errors in metallicity were derived when we merged the catalogs 
(as described in section \ref{sec:feh}), while the errors in the de-aged dust masses were calculated based on the 
photometry errors, with de-aging performed using the flux values at the error boundaries as well. An ideal method 
would include upper limits as censored data, as well as allowing for these sources of error\footnote{Since age errors 
were not readily available, we did not take into account possible errors in the system variables resulting from the 
age uncertainties. Since the errors in metallicity are dominant in the scatter plot, including errors from age 
uncertainties would not have made a significant difference in our conclusions.}. 

We evaluated two linear regression techniques for this purpose, the Gaussian mixed Bayesian (GmB)  method \citep{kelly07} 
and the Akritas-Theil-Sen (ATS) estimator \citep{akritas95}. Datasets with errors in both variables and with censoring 
have been shown to be reasonably well fit with GmB linear regression methods. As a first approach, we fit our data using 
said method with the {\tt linmix\_err} algorithm developed by \cite{kelly07}. Unfortunately, both the metallicity and 
de-aged dust masses have significant heteroscedastic errors and the metallicity vs.\ dust mass plots show large scatter 
with outliers and censoring. The GmB regression method is skewed by outliers, as it assumes a Gaussian error distribution 
for (Y|X). However, our dataset has a non-uniform and non-Gaussian error distribution, and the upper limits in our sample 
do not reflect simple Gaussian errors due to the differing source distances. These issues were apparent upon inspection 
of the GmB fits, so we decided this approach was not applicable in our situation.  

The Theil-Sen \citep{theil50,sen68} regression slope estimator modified by \cite{akritas95}, is able to 
filter outliers and also use censored data. The classic Theil-Sen estimator gives the regression slope as the 
median slope of all pairs of sample points. This algorithm was modified by \cite{akritas95} to allow for censoring. 
To take into account data errors we bootstrapped using the errors in the data, generating 1000 random sets with the 
data distributed based on the parameter errors. We then fitted the distributions with the {\tt cenken} ATS routine in 
{\tt R}, which finds the regression slope via iterations, with the solution being the slope that yields a Kendall $\tau$ of 
0 when removing the values of the test-slope from the data. Therefore, any non-zero regression slope also 
means a non-zero value of the Kendall $\tau$.

\begin{figure*}[t]
\begin{center}
\includegraphics[angle=0,scale=0.55]{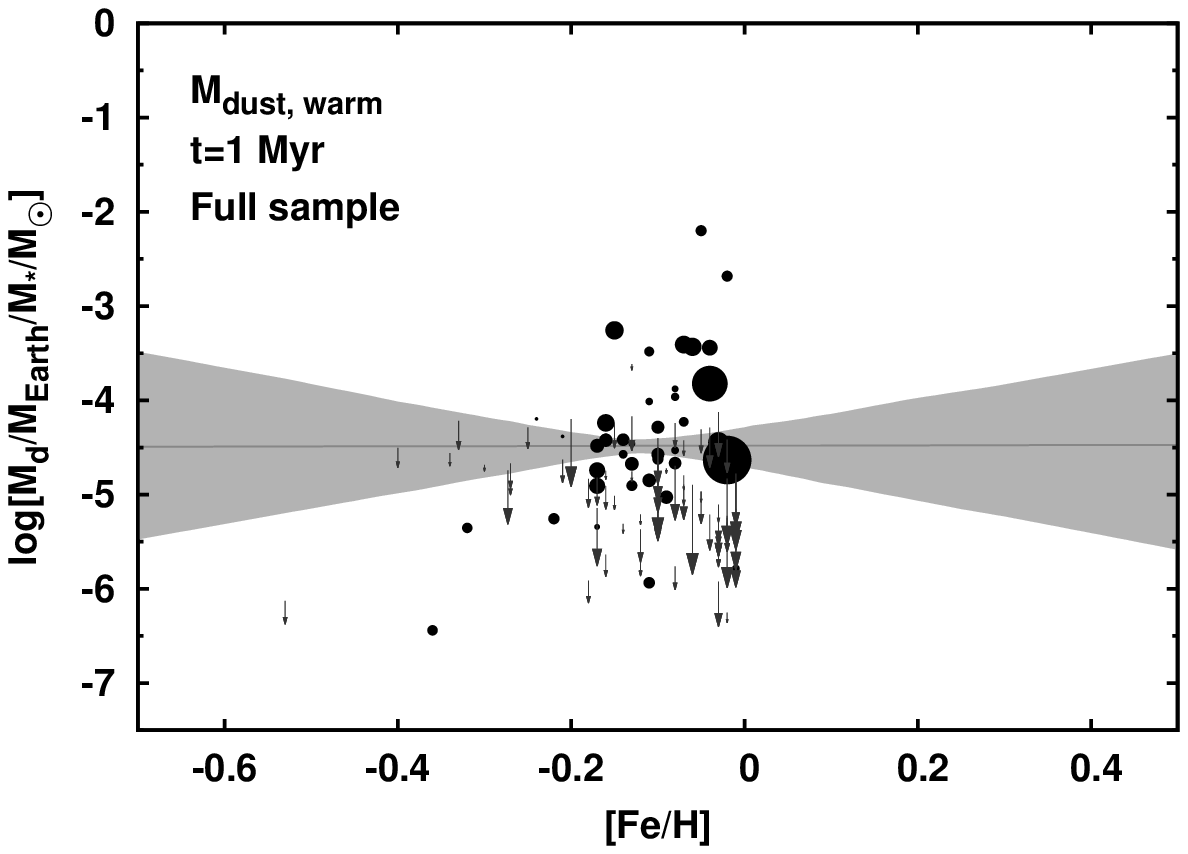}
\includegraphics[angle=0,scale=0.55]{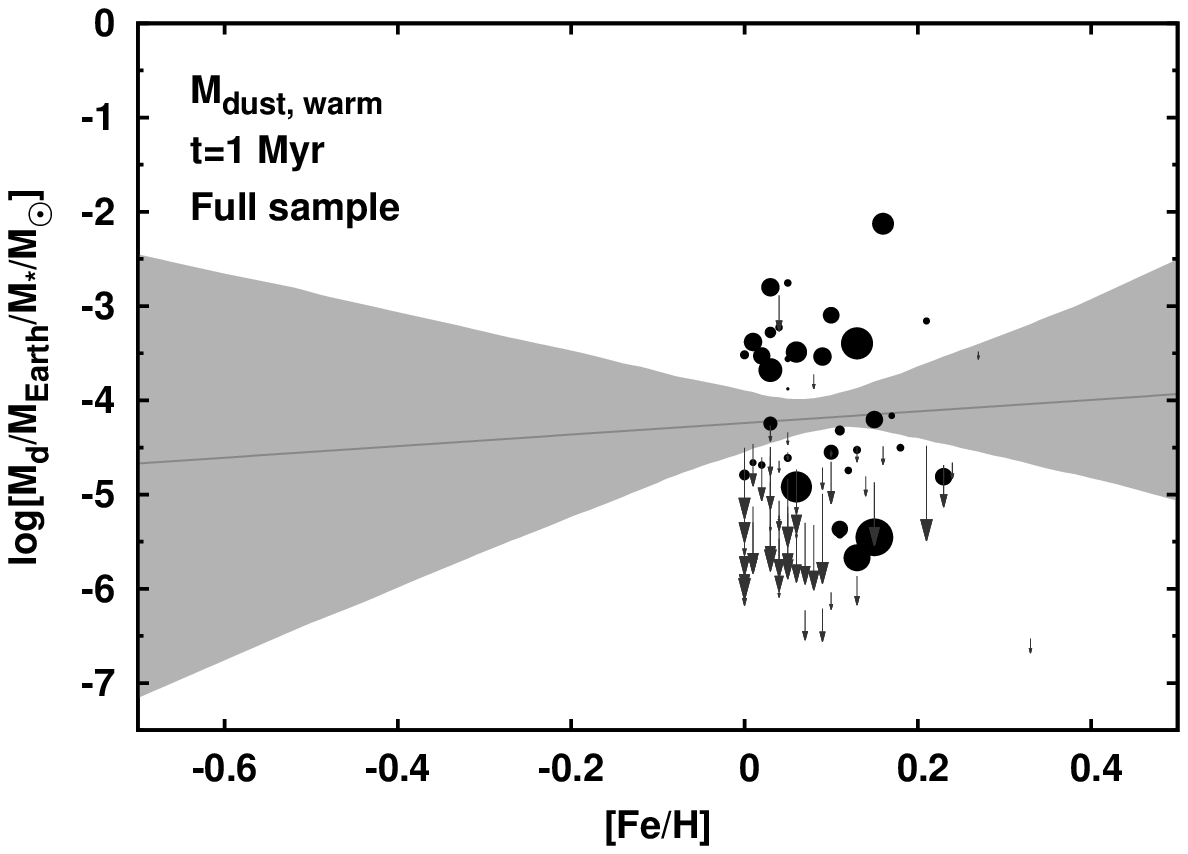}

\includegraphics[angle=0,scale=0.55]{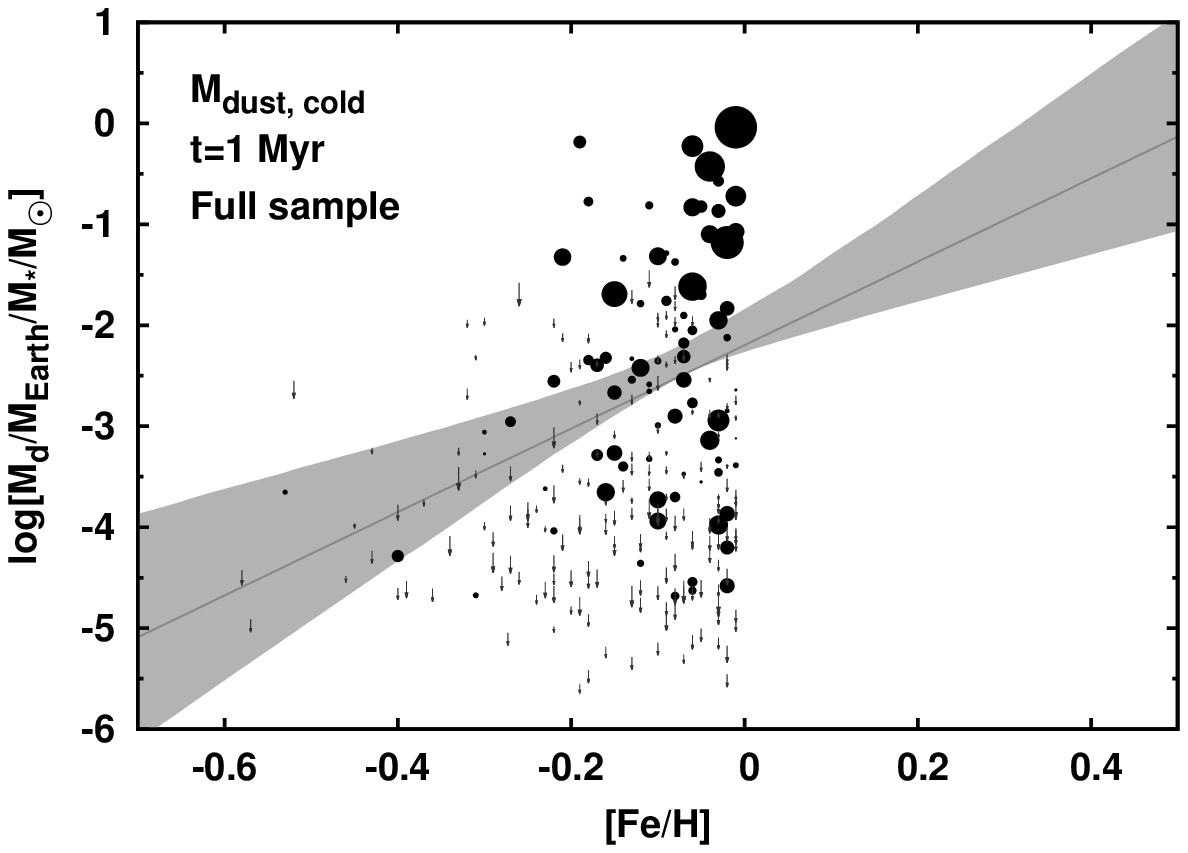}
\includegraphics[angle=0,scale=0.55]{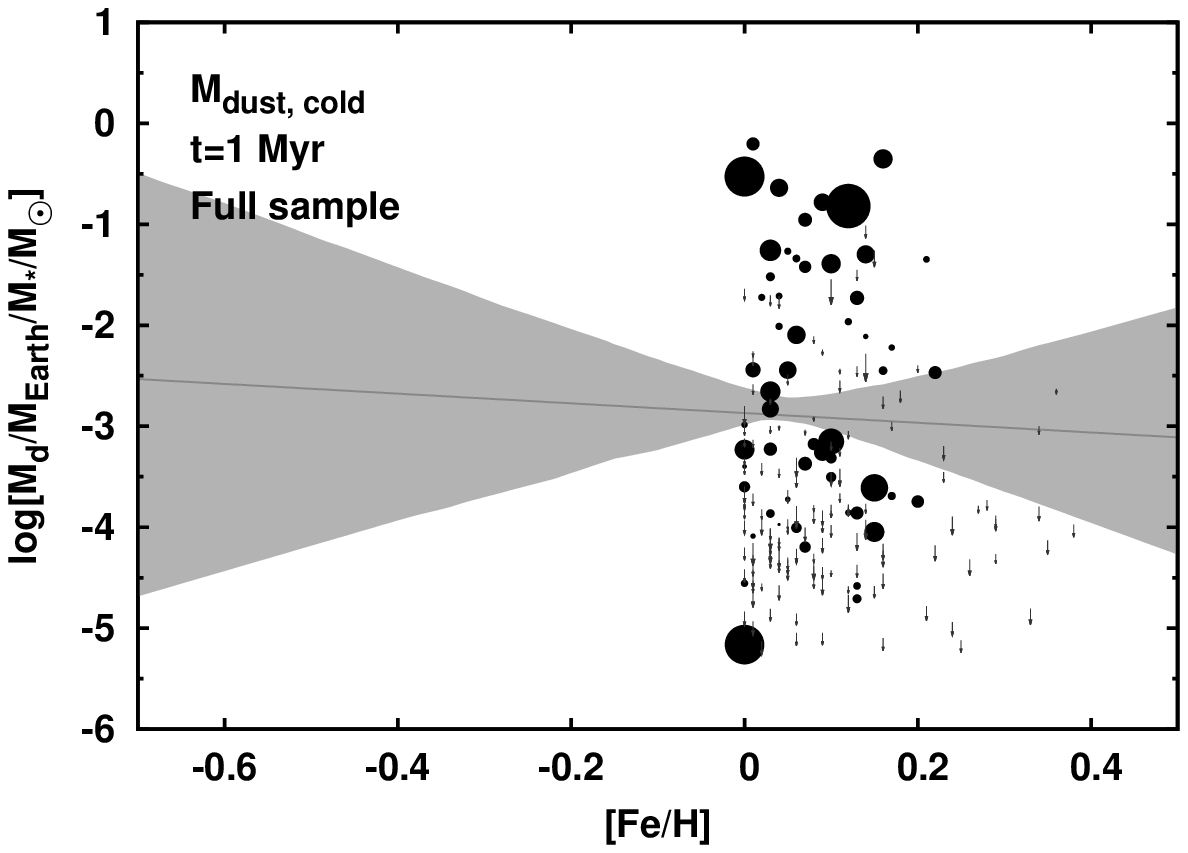}
\caption{Linear regression fits with the warm and cold disk samples divided into low and high metallicity bins at 
${\rm [Fe/H]} = 0$. The best ATS fit is shown as a line, with the 95\% confidence limits in grey. The top row 
is for the warm systems, the bottom row for cold ones. For simplicity we only show the results 
for the full disk sample, de-aged to 1 Myr. The point and arrow 
(upper limits) sizes are inversely proportional to measurement errors, which are generally dominated 
by errors in metallicity. That is, the points with smaller 
errors are plotted with larger symbols.}
\label{fig:coldc}
\end{center}
\end{figure*}

We divided the warm and cold samples at ${\rm [Fe/H]} = 0$ and used the ATS estimator to fit slopes to each metallicity bin. 
The results are in  Figure \ref{fig:coldc}; for simplicity, we only show them for the full spectral sample de-aged to 1 Myr. 
The narrow range in metallicity results in large errors in the slopes. For three cases, warm disks with 
${\rm [Fe/H]} <$ and $> 0$ and cold disks with ${\rm [Fe/H]} > 0$, the slopes are close to zero but the 
95\% probability error ranges allow a correlation as strong as disk mass going as the square of the metallicity. 
For the cold disks and ${\rm [Fe/H]} < 0$, a significant slope is indicated in the expected direction, i.e., 
disks of significant mass are uncommon around low metallicity stars. 

\section{Conclusions}
\label{sec:conclude}

Since their early discovery \citep{aumann84,smith84,backman93}, debris disks have always been thought of 
as signposts of planetary systems. However, there are few examples where direct connections between debris 
disk properties and those of exoplanetary systems have been established. In this paper we revisited 
this conundrum.

We gathered {\it Spitzer} and {\it Herschel} data on 199 warm disks (of which 125 only have
upper limits) and 463 cold disks (of which 315 only have upper limits) around 478 hosts. 
Through the merging of 33 individual spectral catalogs, we also collected metallicity data for these
systems. Using our collisional cascade model \citep{gaspar12a}, we then ``de-aged'' the 
systems, yielding dust masses for the systems at the time of their formation. We found multiple indications 
that massive debris disks are uncommon at low stellar metallicity (e.g., ${\rm [Fe/H]} < -0.2$. Although not 
all of these indicators individually are at a high level of statistical significance, their combination points 
to a real avoidance by debris disks of low metallicity environments. 

These results are in qualitative agreement with the planet-metallicity correlation \citep{gonzalez97,fischer05} 
and the core-accretion formation model \citep{mizuno80,ikoma00} (where the heavier elements aggregate into dust
and further on to planetesimals) and the type II migration models of close orbit giant planets 
\citep{goldreich80,armitage07}. \cite{fischer05} found that the planet-metallicity correlation is 
a result of system initial conditions rather than acquired via accretion, i.e.\ planetary 
systems with close orbit giant planets are more likely to form from molecular clouds with higher initial metal content. 
Debris disks generally bound the regions of giant planet formation, with the warm components 
defining the inner and the cold components the outer edges. Our finding therefore corresponds roughly with the 
strong dependence of the incidence of giant planets on metallicity \citep[e.g.,][]{fischer05,johnson10, wang15a}.

\acknowledgements

Support for this work was provided by NASA through Contract Number 1255094 
issued by JPL/Caltech. This research has made use of the SIMBAD database, 
operated at CDS, Strasbourg, France.

\end{document}